\documentclass[twocolumn,amsmath,amssymb,aps,nofootinbib,superscriptaddress]{revtex4-1}
\usepackage{iftex}
\ifXeTeX
\usepackage{fontspec}
\fi
\usepackage{subcaption}
\usepackage{graphicx}
\usepackage{caption}
\usepackage{amsmath}
\usepackage{hyperref}
\usepackage{color}

\renewcommand{\eqref}[1]{eq.~\ref{#1}}

\definecolor{alizarin}{RGB}{231, 76, 60}

\graphicspath{{./}{img/}}

\begin{document}

\begin{flushright}
\hspace{12cm} INT-PUB-19-005
\end{flushright}

\title{Roper State from Overlap Fermions}

\author{Mingyang Sun}
\affiliation{\mbox{Department of Physics and Astronomy, University of Kentucky, Lexington, Kentucky 40506, USA}}

\author{Ying Chen}
\affiliation{\mbox{Institute of High Energy Physics, Chinese Academy of Sciences, Beijing 100049, China}}

\author{Gen Wang}
\affiliation{\mbox{Department of Physics and Astronomy, University of Kentucky, Lexington, Kentucky 40506, USA}}

\author{Andrei Alexandru}
\affiliation{\mbox{Department of Physics, The George Washington University, Washington, D.C. 20052, USA}}

\author{Shao-Jing Dong}
\affiliation{\mbox{Department of Physics and Astronomy, University of Kentucky, Lexington, Kentucky 40506, USA}}

\author{Terrence Draper}
\affiliation{\mbox{Department of Physics and Astronomy, University of Kentucky, Lexington, Kentucky 40506, USA}}

\author{Jacob Fallica}
\affiliation{\mbox{Department of Physics and Astronomy, University of Kentucky, Lexington, Kentucky 40506, USA}}

\author{Ming Gong}
\affiliation{\mbox{Institute of High Energy Physics, Chinese Academy of Sciences, Beijing 100049, China}}

\author{Frank X. Lee}
\affiliation{\mbox{Department of Physics, The George Washington University, Washington, D.C. 20052, USA}}

\author{Anyi Li}
\affiliation{\mbox{Institute for Nuclear Theory, University of Washington, Seattle, 98195, USA}}

\author{Jian Liang}
\affiliation{\mbox{Department of Physics and Astronomy, University of Kentucky, Lexington, Kentucky 40506, USA}}

\author{Keh-Fei Liu}
\email{liu@g.uky.edu}
\affiliation{\mbox{Department of Physics and Astronomy, University of Kentucky, Lexington, Kentucky 40506, USA}}

\author{Nilmani Mathur}
\affiliation{\mbox{Department of Theoretical Physics, Tata Institute of Fundamental Research, Mumbai 400005, India}}

\author{Yi-Bo Yang}
\affiliation{\mbox{Institute of Theoretical Physics, Chinese Academy of Sciences, Beijing 100190, China}}


%
%
%
%
%

\begin{abstract}

\begin{center}
\large{
\vspace*{0.0cm}
\includegraphics[scale=0.20]{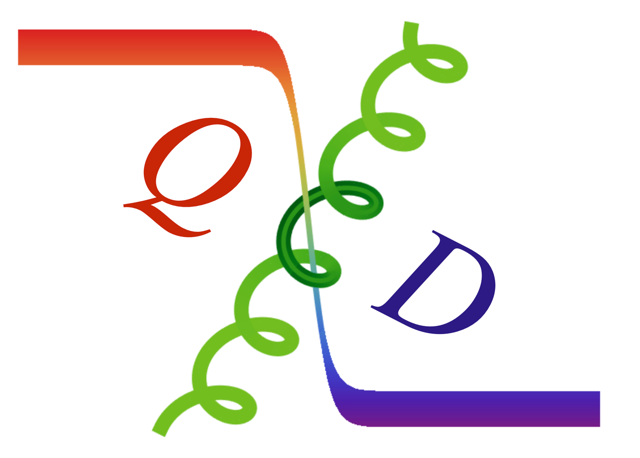}\\
\vspace*{0.4cm}
($\chi$QCD Collaboration)
\vspace*{0.4cm}
}
\end{center}

The Roper state is extracted with valence overlap fermions on a $2+1$-flavor
domain-wall fermion lattice (spacing $a = 0.114$ fm and $m_{\pi} = 330$ MeV)
using both the Sequential Empirical Bayes (SEB) method and the variational
method. The results are consistent, provided that a large smearing-size
interpolation operator is included in the variational calculation to have
better overlap with the lowest radial excitation. The SEB and variational
calculation with large smearing size are also carried out for an anisotropic clover lattice with
similar parameters (spatial lattice spacing $a_s = 0.12$ fm and pion mass
$m_{\pi} = 396$ MeV) and obtain consistent results. 
However, these calculations with clover fermions 
give a Roper mass of $m_R = 1.92(6)$ GeV, while the same approach with overlap
fermions finds the Roper $\approx 280$ MeV lower, at $m_R = 1.64(9)$ GeV, for
identical valence pion mass. The fact that the prediction of the Roper state by
overlap fermions is consistently lower than those of clover fermions, chirally
improved fermions, and twisted-mass fermions over a wide range of pion masses
has been dubbed a ``Roper puzzle.''

To understand the origin of this difference, we study the hairpin $Z$-diagram in
the isovector scalar meson ($a_0$) correlator in the quenched
approximation. The lack of quark loops in the quenched approximation turns the
$a_0$ correlator negative; giving rise to a ghost ``would-be'' $\eta\pi$ state.
Comparing the $a_0$ correlators for valence clover and overlap fermions,
at a valence pion mass of 290 MeV, on three quenched Wilson-gauge lattices,
we find that the spectral weight of the ghost state with clover fermions is
smaller than that of the overlap at $a = 0.12$ fm and $0.09$ fm $--$  the ratios of the Wilson
ghost-state magnitudes (correlator minima) are about half of those of overlap $--$
whereas, the whole $a_0$ correlators of clover and overlap at $a = 0.06$ fm coincide within
errors. This suggests that chiral symmetry is restored for clover at $a \le 0.06$ fm and
that the Roper mass should agree between clover and overlap fermions toward the 
continuum limit.

We conclude that the present work supports a resolution of the ``Roper puzzle''
due to $Z$-graph type chiral dynamics. This entails coupling to higher
components in the Fock space (e.g.  $N\pi$, $N\pi\pi$ states) to induce the
effective flavor-spin interaction between quarks as prescribed in the chiral
quark model, resulting in the parity-reversal pattern as observed in the experimental
excited states of $N, \Delta$ and $\Lambda$. \\

\end{abstract}

\maketitle

\section{I. Introduction}    \label{introduction}

The nature of the lowest nucleon excited state, the Roper N(1440) which appears
in the $\pi N$ scattering in the $P_{11} $ channel at about 1370 MeV with a
width of about 175 MeV, has been a controversial and intriguing subject since
its discovery.  First of all, it is rather unusual to have the first
positive-parity excited state lower than the negative-parity excited state,
which is the $N^{\frac{1}{2}^-}(1535)$ at 1510 MeV in the $\pi N\,S_{11}$
scattering channel.  This is contrary to the excitation pattern in the meson
sectors with either light or heavy quarks.  This parity reversal has caused
problems for the otherwise successful quark models based on $SU(6)$ symmetry
with color-spin interaction between the quarks, which cannot accommodate such a
pattern.  Realistic potential calculations with linear and Coulomb
potentials~\cite{liu_cluster_1983} and the relativistic quark
model~\cite{capstick_baryons_1986,capstick_quark_2000} all predict the Roper to
be $\sim 100$--200 MeV above the experimental value, and above the negative
parity state.  On the other hand, the pattern of parity reversal was readily
obtained in a chiral soliton model, the Skyrme model, via the small oscillation
approximation to $\pi N$ scattering.  Although the first
calculations~\cite{liu_time_1984, breit_phase_1984} of the original Skyrme
model gave rise to a breathing mode which is $\sim 200\,{\rm MeV}$ lower than
the Roper resonance, it was shown later~\cite{kaulfuss_breathing_1985} that the
introduction of the sixth-order term, which is the zero-range approximation for
the $\omega$ meson coupling, changes the compression modulus and yields better
agreement with experiment for both the mass and width in $\pi N$ scattering.

Since the quark potential model is based on $SU(6)$ symmetry with residual
color-spin interaction between the quarks, whereas the chiral soliton model is
based on spontaneous broken chiral symmetry, their distinct predictions on the
ordering of the positive- and negative-parity excited states are most likely a
reflection of different dynamics derived from their respective symmetries.
This possibility has prompted the suggestion~\cite{glozman_spectrum_1996} that
the parity reversal in the excited nucleon and $\Delta$, in contrast to that in
the excited $\Lambda$ spectrum, is an indication that the inter-quark
interaction of the light quarks is mainly of the flavor-spin nature rather than
the color-spin nature (e.g.\ one-gluon exchange type). This suggestion is
supported in the lattice QCD study of ``Valence QCD''~\cite{Liu:1998um} which
finds that the hyperfine splitting between the nucleon and $\Delta$ is greatly
diminished when the $Z$-graphs in the quark propagator are eliminated by
modifying the fermion action to prevent the quarks from propagating backwards
in time.  This is an indication that the color-magnetic interaction is not the
primary source of the inter-quark spin-spin interaction for light quarks. (The
color-magnetic part, being spatial in origin, is unaffected by the truncation
of $Z$-graphs in Valence QCD, which only affects the temporal part.)  Yet, it is
consistent with the flavor-spin interaction being generated by the quark
$Z$-graph in the Goldstone-boson-exchange picture. This picture is enhanced by
the extensive dynamical coupled- channel (DCC) model analysis carried out by
the Excited Baryon Analysis Center [EBAC] at
JLab~\cite{JuliaDiaz:2007kz,Kamano:2010ud,Suzuki:2009nj}. This DCC Hamiltonian
approach involves a nucleon core and the meson-baryon reactions in the $\pi N$,
$\eta N$, and $\pi\pi N$ channels which fits 22,348 independent data points,
representing the complete array of partial waves below 2 GeV. In the $\pi N$
$P_{11}$ partial wave, it is found that starting with a bare state at 1.763
GeV, two resonances merge into the Roper resonance. They are at 1357 MeV
with $\Gamma = 152$ MeV and 1364 MeV with $\Gamma = 210$ MeV. There is a third
resonance at 1820 MeV with $\Gamma = 496$ MeV which may correspond to
$N^*(1710)$. It is interesting to note that the meson-baryon coupling brings
down the bare state by $\approx 400$ MeV, to the Roper mass. The failure of
the $SU(6)$ quark model to delineate the Roper and its photo-production has
also prompted the speculation that the Roper resonance may be a hybrid state
with excited glue~\cite{barnes_where_1983,Li:1991sh,Carlson:1991tg} or a
$qqqq\bar{q}$ five-quark state~\cite{Krehl:1999km,Jaffe:2004zg}.  Thus,
unraveling the nature of Roper resonance has direct bearing on our
understanding of the quark structure and chiral dynamics of baryons, which is
one of the primary missions at experimental facilities such as at Jefferson
Lab~\cite{Burkert:2017djo}. At the moment, the nature of Roper and why it is
lower than the quark model prediction as the radial excitation of the nucleon
is unsettled.

Lattice QCD, being a first-principles approach, is regarded as the most
desirable tool to adjudicate the theoretical controversy surrounding the issue
and to discern the nature of Roper. However, lattice calculations of the Roper
state are also shrouded in a puzzle which we shall address and sort out in this
manuscript. Studying the ghost would-be $\eta\pi$ state in the quenched
approximation has paved the way to better understand the chiral dynamics and
the origin of the difference due to fermion actions with and without chiral
symmetry at finite lattice spacing.

This manuscript is organized as follows. The status of lattice calculations of
Roper are reported in Sec. II where the discrepancy between results from
Wilson-type fermions and those from the overlap fermions is pointed out.
Sec. III will see calculations of both the SEB and variational methods for the
clover and overlap fermions to check the consistency of the two approaches in
obtaining the nucleon excited states.  In an attempt to understand the puzzling
difference of the Roper between the two fermion actions in terms of chiral
dynamics, we calculate the ghost would-be-$\eta\pi$ state in the quenched
approximation for the overlap and Wilson fermions in Sec. IV. This is a
sensible and pertinent place to compare the coupling strength of single-hadron
interpolating fields to two-hadron states for these fermion actions as a
function of lattice spacing. We finish in Sec. V with a summary and conclusion.

\begin{figure}[t]
    \includegraphics[width=1.0\linewidth]{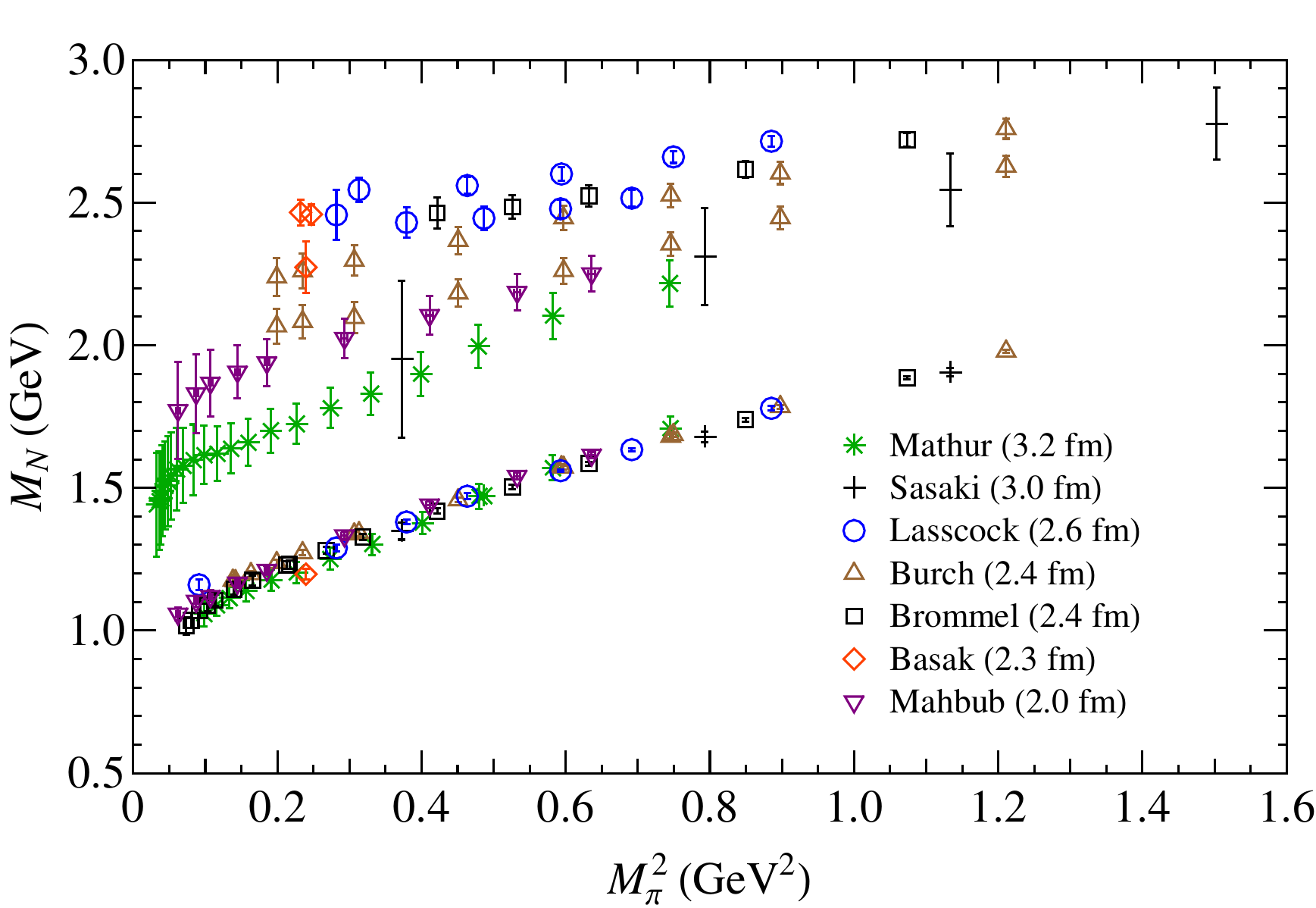}
    \caption{Quenched results of masses of the nucleon ground state and first
      positive-parity excited state for a broad range of pion masses.  The
      green stars are results obtained from the overlap fermion.}
    \label{fig:roper-quenched}
\end{figure}

\begin{figure*}
    \includegraphics[width=1.0\linewidth]{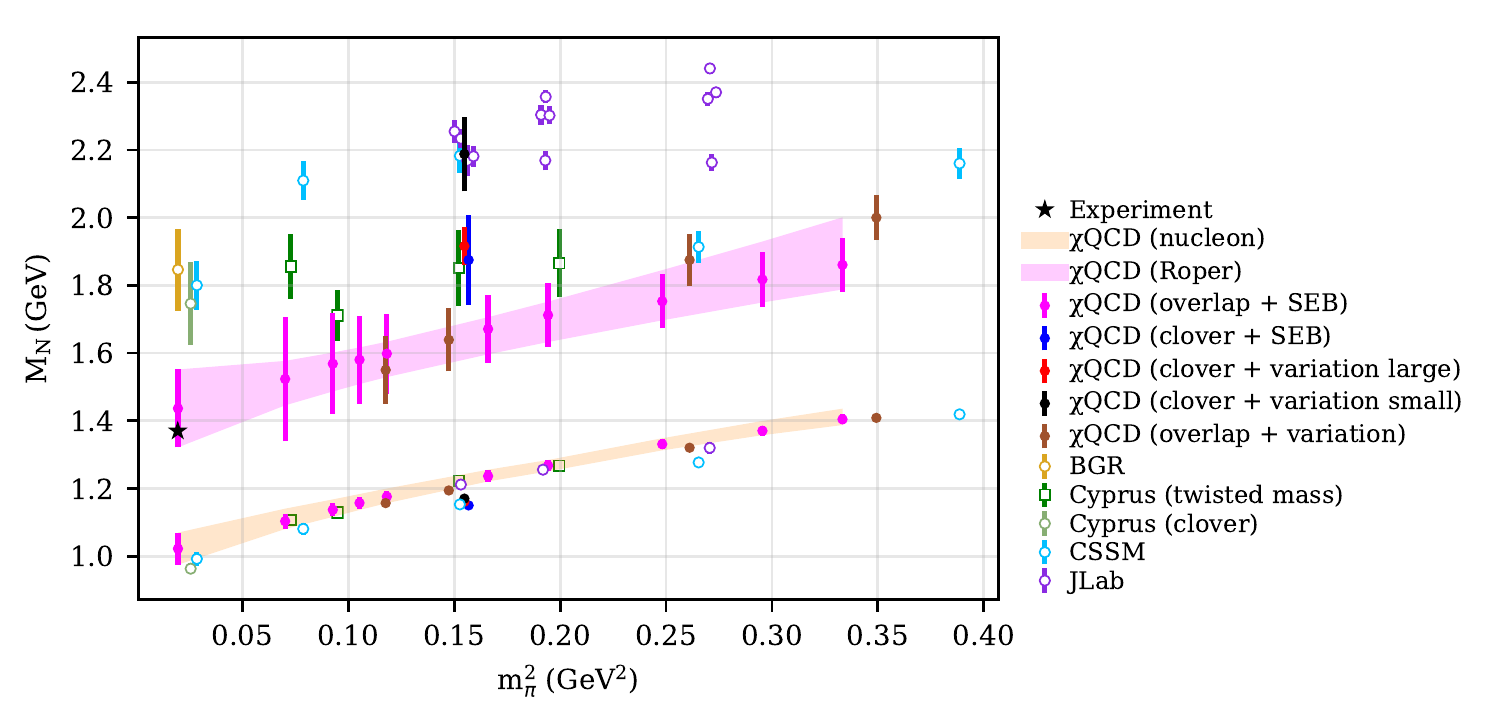}
    \caption{Results with dynamical fermion configurations for the masses of
      the nucleon ground state and its first positive-parity excited state
      across several pion masses.  The magenta points and band are results
      obtained from valence overlap fermions on the domain-wall sea with the SEB method.
      The brown points are from the variational results on the same lattice. The blue point is the
      SEB result on the JLab clover configurations. The red/black point is the variational result
      on the JLab clover configurations with large/small interpolation fields. Also listed are
      results from BGR Collaboration~\cite{Engel:2013ig}, Cyprus with twisted mass and 
      clover fermions~\cite{Alexandrou:2013fsu}, CSSM with clover fermions ~\cite{mahbub_roper_2012,mahbub_structure_2013},
      and JLab with clover fermions on an anisotropic lattice~\cite{Edwards:2011jj}.       }
  \label{fig:roper-dynamical}
\end{figure*}

\section{II. Roper puzzle from lattice calculations}    \label{Roper-lattice}

Lattice calculations of the positive-parity excitation of the nucleon started
out with the quenched approximation. The first set of calculations used a
nucleon interpolation field which does not have a non-relativistic limit to
project to the excited $1/2^+$
state~\cite{Lee:2000hha,Sasaki:2001nf,Richards:2001bx,Melnitchouk:2002eg,Edwards:2003cd}
and found it to be much higher than the Roper. Later, Bayesian
methods~\cite{Sasaki:2005ap,mathur_roper_2005} and variational approaches with
multiple interpolation
operators~\cite{Brommel:2003jm,Basak:2006ww,Burch:2006cc,Lasscock:2007ce,Mahbub:2010jz}
were introduced in the calculation of the excited state. The results of the
calculations are plotted in Fig.~\ref{fig:roper-quenched} as a function of
$m_{\pi}^2$ together with the corresponding nucleon masses. The nucleon masses
from different calculations are all in agreement within errors.  However, one
distinct feature stands out, i.e., the excited $\frac{1}{2}^+$ states do not
approach the experimental Roper state at the physical pion point upon chiral
extrapolation. The only exception is overlap fermions for which the Roper mass
is consistently lower than in the other calculations at each value of
$m_{\pi}^2$. The situation with the dynamical fermion calculations is basically
a replication of that of the quenched approximation. As shown in
Fig.~\ref{fig:roper-dynamical}, the variational calculations of $N_f = 2+1$
clover fermions from the CSSM
Collaboration~\cite{mahbub_roper_2012,mahbub_structure_2013}, an anisotropic
lattice calculation from the JLab Collaboration~\cite{Edwards:2011jj}, an
$N_f = 2$ clover and twisted-mass fermions studies from the Cypress
group~\cite{Alexandrou:2013fsu}, and a chirally improved fermion study from the
BGR Collaboration ({\it N.B.} The result is from chiral extrapolation to the
physical pion mass)~\cite{Engel:2013ig}, all yield much higher Roper masses at
the same $m_{\pi}^2$ than those of the overlap
calculation~\cite{Liu:2014jua,Liu:2016rwa} which uses the SEB
method. Especially near the physical pion mass, the results from the clover
fermion~\cite{mahbub_structure_2013,Alexandrou:2013fsu}, are $\sim$400 MeV
above the experimental Roper mass at 1370 MeV. The twisted mass calculation
from the Cypress group ~\cite{Alexandrou:2013fsu}, with large smearing sources
in the variational approach, gives lower Roper masses than those of the clover
fermion and is closer to that of the overlap fermion at $m_{\pi} \sim 300$ MeV,
but higher at other pion masses, especially at $m_{\pi} = 270$ MeV.
Recently, a calculation was made that used both $q^3$ and $q^4\bar{q}$ (interpolation
field in the form of $N \pi$ and $N\sigma$) operators on a $2+1$-flavor clover
fermion lattice with $a = 0.0907$ fm and $m_{\pi} = 156$ MeV to check if the
Roper state could be produced through the coupling of higher Fock
space~\cite{Lang:2016hnn}. No Roper state was found below 1.65 GeV in this
coupled-channel scattering calculation. The same conclusion has been drawn from
a similar calculation on the same clover fermion lattice except at $m_{\pi} =
411$ MeV~\cite{Kiratidis:2017bnr}.

In contrast to the clover and twisted-mass results, it has been
shown~\cite{Liu:2014jua,Liu:2016rwa} in Fig.~\ref{fig:roper-dynamical} that
the first positive-parity excited nucleon state from overlap fermions with the
SEB method is consistently $\sim 500$ MeV above the corresponding nucleon mass
in the pion mass range up to $\sim 580$ MeV. This behavior is consistent with
the typical size of the radial excitation of $\sim$500 MeV for the nucleon,
$\Delta, \Lambda$ and $\Sigma$ as well as heavy quarkoniums. The overlap
calculation with the SEB method is based on $2+1$-flavor domain-wall fermion
configurations on the $24^3 \times 64$ lattice with $a = 0.114$ fm and sea pion
mass at 330 MeV. Multiple overlap valence quark masses are used in this
partially-quenched calculation to cover the valence pion mass range from 260
MeV to 580 MeV. We fit the available data for different quark masses with the
form
\begin{equation}
M_{N,R} = M_{N,R}(0)+ c_{2 (N,R)} m_{\pi, vv}^2 + c_{3 (N,R)} m_{\pi, vs}^3
\end{equation}
{where $m_{\pi,vv}$ is the valence pion mass. The $m_{\pi,vs}^3$ term
is introduced to reflect the effects of partial quenching between the valence and the sea quark mass. The partially quenched pion mass $m_{\pi,vs}$ is made of a combination of the valence quark/antiquark and the sea antiquark/quark with the relation $m_{\pi, vs}^2 =1/2(m_{\pi,vv}^2 + m_{\pi,ss}^2) + \Delta_{mix} a^2$, where $\Delta_{mix}$ is a low-energy constant for the mixed action. Since the overlap fermion action and the domain wall fermion action are close to each other, we obtain a small $\Delta_{mix}$~\cite{Lujan:2012wg}. The value 
$\Delta_{mix} = 0.030(6)(5)\, {\rm GeV}^4$ we obtained is calculated on lattices with two lattice spacings (0.014 fm and 0.085 fm) and is an order of magnitude smaller than those of DWF on staggered and overlap on clover~\cite{Lujan:2012wg}. It gives a shift of only 16 MeV to the valence pion mass at 300 MeV for the lattice we use in this work.}  At the physical pion limit, we obtain $M_N = 0.999(46)$ GeV and 
 $M_R = 1.40(11)$ GeV, in good agreement with experiment.
 {In Fig.~\ref{fig:roper-dynamical}, the data points are from the SEB fit to the results from the full gauge ensemble and the error bars are obtained directly from the fit. We also carry out a single-elimination jackknife analysis by performing the SEB fit to each jackknive ensemble, and treat the jackknife error of each data point as the statistical error to build the covariance error matrix. Based on this jackknife analysis, the chiral extrapolation was carried out using Eq. (1) in the manuscript, which is illustrated in the figure by the error band.}

Based on this observation, it has been speculated that the difference could be
due to the fact that the overlap fermion is a chiral fermion which induces
different dynamics for the excited nucleon.  However, there is a caveat in that
the overlap calculation employed the SEB method~\cite{Chen:2004gp}. This
method, although successfully applied in calculating
$S_{11}(1535)$~\cite{mathur_roper_2005}, $a_0(1450)$ and
$\sigma(600)$~\cite{mathur_lattice_2007}, and the radially excited $1P$
charmonium state~\cite{Chen:2007vu}, has not been tested by other groups. The
variational approach is considered a robust and trustworthy algorithm for
excited-state calculation. Thus, there is no consensus on this puzzle and one
needs to sort out the issue regarding different algorithms first.

\section{III. Variational and Sequential Empirical Bayes calculations}  \label{variation-clover}

The only sensible way to check if the discrepancy between the overlap fermion
results from the SEB method and those of other fermions from the variational
method is due to the different algorithmic approaches is to apply both methods
for the same fermion action and on the same lattice to see if they produce
consistent results. To this end, we choose to examine two cases: an ensemble
using clover fermions from JLab~\cite{Dudek:2012gj,Jung:2017xef} and another
using valence overlap fermions on a domain-wall sea generated by
RBC/UKQCD~\cite{Blum:2014tka,Jung:2017xef}. The parameters of the
clover~\cite{Dudek:2012gj} and the domain-wall
~\cite{Blum:2014tka,Jung:2017xef} lattices are tabulated in
Table~\ref{lattices}.

We first applied the SEB method on the clover configurations with the same
clover fermion as used by the JLab group, except that we worked with a larger
$24^3 \times 128$ lattice instead of the $16^3 \times 128$ lattice used to
calculate Roper in Ref.~\cite{Edwards:2011jj} at the same quark mass. The
result is plotted in Fig.~\ref{fig:roper-dynamical}. We see that the nucleon
mass (blue point) is consistent with that from the JLab
calculation~\cite{Edwards:2011jj}, but the first positive-parity state
from SEB (blue point) at $1.87(13)$ GeV is lower than that of the JLab
calculation at $\sim 2.20(10)$ GeV by $\sim$300 MeV with more than a 2.5
$\sigma$ difference. To check if this difference is due to the fitting
algorithms, we performed variational calculations with different
gauge-invariant Gaussian smearing sizes for the nucleon source and sink.

\begin{table}[htbp]
\begin{center}  
\caption{\label{table:lattice} Lattice parameters of the clover and domain-wall
  fermion configurations including the lattice size, spatial lattice spacing
  $a_s$, the anisotropic factor $\xi$, the pion mass from the light sea quark
  mass, and the number of gauge configurations $N_{\rm conf}$ are listed.}
\begin{tabular}{ccccccc}     \label{lattices}
Fermion & $L^3\times T$  &$a_s$ (fm)  & $\xi$ & $L (\rm fm)$&  $m_{\pi}$ (MeV) & $N_{\rm conf}$  \\
\hline
Clover & $24^3\times 128$& 0.123 & 3.5 &2.95  &396   & 761  \\
Domain Wall &$24^3\times 64$& 0.114 &  1  &2.74  &330   & 203\\
\hline
\end{tabular}
\end{center}
\end{table}

\begin{figure}[h]  
\begin{minipage}{0.5\textwidth}
  {\includegraphics[width=0.84\hsize,height=0.65\textwidth]{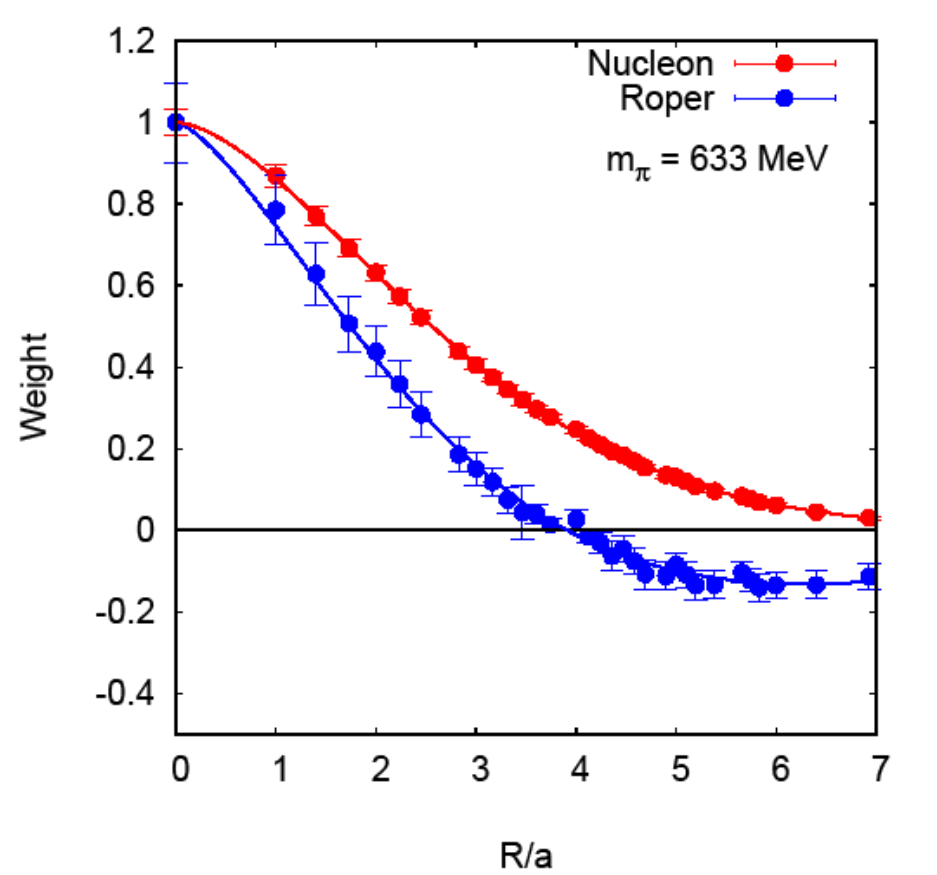}}
\end{minipage}
\begin{minipage}{0.5\textwidth}
\vspace*{-2.2cm}
\hspace*{-0.7cm}
  {\includegraphics[width=0.95\hsize,height=1.0\textwidth]{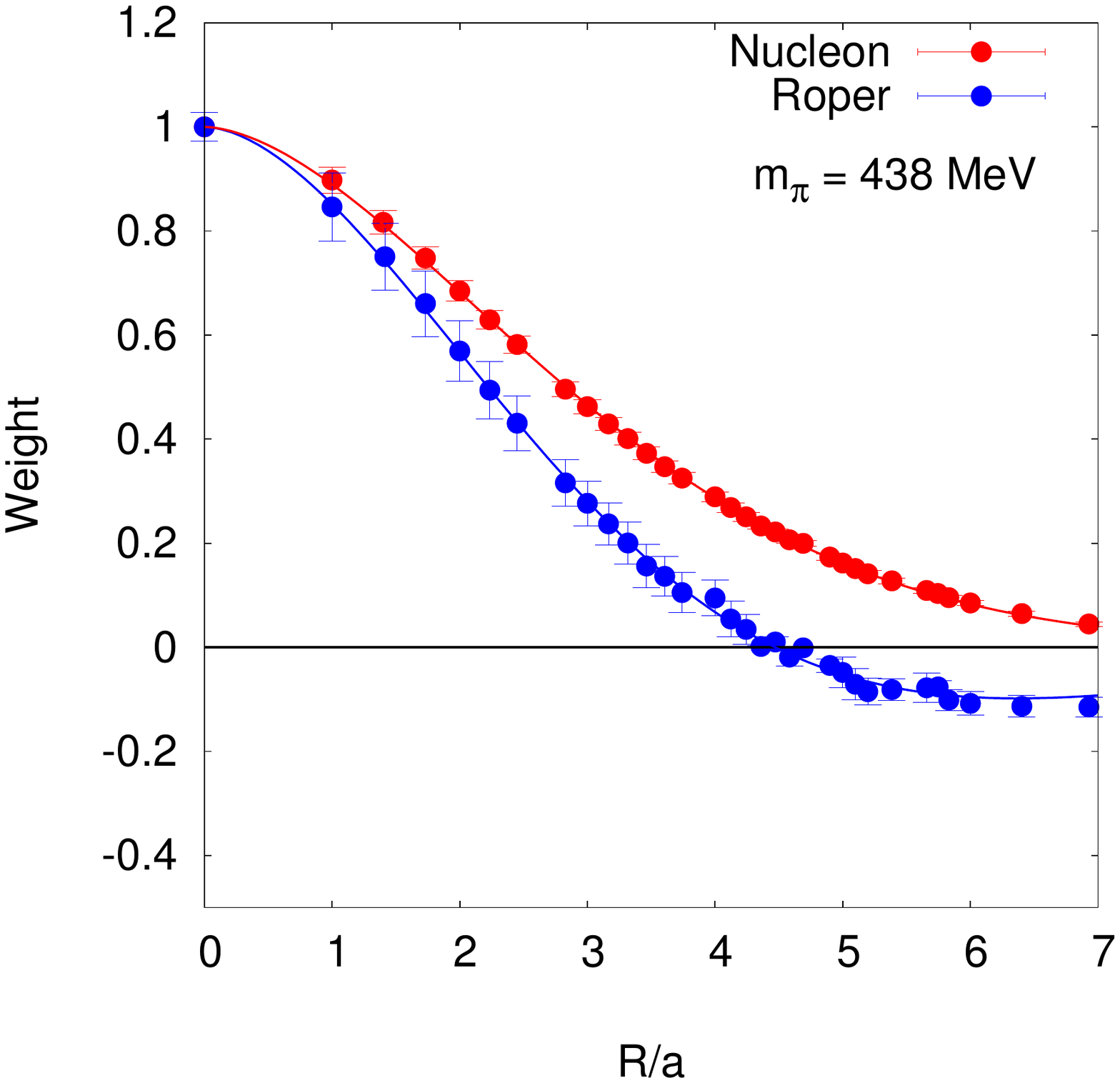}}
 \end{minipage}
 \vspace*{-0.5cm}
  \caption{Nucleon and Roper wavefunctions in the Coulomb gauge for two different pion masses.}
  \label{Roper_WF}
\end{figure}

Before addressing the variational calculation, it is worthwhile noticing that,
to the extent that the Roper is the radially excited nucleon, its primary
distinction from the nucleon is its radial wavefunction. This will affect the
smearing size of the source and sink to be used in the variational
calculation. To this end, we show in Fig.~\ref{Roper_WF} the Roper and nucleon
wavefunctions which were obtained by placing two $u$ quarks of the sink
interpolation field for the proton at the same point and the $d$ quark at a distance $R$ away
on a Coulomb gauge-fixed lattice. These were obtained from a quenched
calculation with the overlap fermion on a $16^3 \times 28$ lattice with $a =
0.2$ fm~\cite{chen_coulomb_2007}.  We note that for $m_{\pi}$ at both 633 MeV
and 438 MeV, there is a node at $R = 0.76$ and 0.9 fm respectively, indicating
that it is a radial excitation of the nucleon with perhaps some higher Fock
space components, such as $N\pi$, $N\eta$ and $N\pi\pi$ states. A variational
calculation with dynamical clover fermions at $m_{\pi} = 156 $ MeV has also
observed a node at $R \sim 0.8$ fm in the Roper
wavefunction~\cite{Roberts:2013oea} in Landau gauge.
{We note that each Roper wavefunction in Fig.~\ref{Roper_WF} does not damp off at R = L/2 = 1.6 fm in Fig.~\ref{Roper_WF}  which indicates that there will be certain finite size effect for the wavefunction. As far as the spectrum is concerned, however, $m_{\pi} L = 10$ for $m_{\pi} = 633$ MeV (upper panel) and 7 for $m_{\pi} = 438$ MeV (lower panel). Thus, we expect that the Roper mass is not much affected. The purpose of Fig.~\ref{Roper_WF} is to reveal that the Roper is mainly a radial excitation with a node in its radial wavefunction. The finite size effect of the wavefunction is not a primary concern in this study.}

The existence of a node in the Roper wavefunction explains the observation that
the spectral weight for the Roper from the Coulomb wall source and point sink
is negative, in contrast to that of the nucleon.  Since the spectral weight of
the Roper with the Coulomb wall source is the product of two factors, one of
which is the positive matrix element $\langle 0|\sum_x \chi (x,x,x)|{\rm
  Roper}\rangle$ for the sink, the sign of the wall-source matrix element
$\langle {\rm Roper}|\sum_{x,y,z} \chi (x,y,z)|0\rangle$ controls the overall
sign. The extra spatial sums in this matrix element involve an $R^2$ measure
which gives more weight to positions with larger $R$, so the node in the Roper
wavefunction (seen in Fig.~\ref{Roper_WF}) turns the overall weight
negative. Thus, both the explicit calculation of the radial wavefunction and
the negative spectral weight lend concrete evidence to the fact that the Roper
is primarily a radial excitation of the nucleon with perhaps some higher
Fock-space components.  These lattice results have ruled out that the Roper is
a pure pentaquark state, since such a state would have one quark or anti-quark
in the relative P-wave and the rest in the lowest S-waves. This would not lead
to a radial node in the Bethe-Salpeter wavefunction, in contrast to
observation.

\begin{table}[b]
\caption{Values used for the smearing parameter $w$ for the clover and overlap
  variational calculations, the number of iterations used for the Gaussian
  smearing, and the measured root-mean-square (rms) radius.}
\label{tab:smearing}
 \begin{tabular}{|c|cc|c|cc|}
 \hline
 \multicolumn{3}{|c} {clover} & \multicolumn{3}{|c|} {overlap}  \\
 \hline
 \multicolumn{1}{|c|}{$w$} & \multicolumn{1}{c}{iterations} & $\langle r^2\rangle^{1/2}$(fm)  & \multicolumn{1}{|c|}{$w$} & iterations & 
 $\langle r^2\rangle^{1/2}$(fm) \\
 \hline
     0 & 0 & 0 & 0 & 0 &0 \\
    2 & 50 & 0.19 & 2 & 50  & 0.16\\
    4 & 100 & 0.39 & 4& 100 & 0.34 \\
    7 & 200 & 0.63 & 8 & 100 & 0.63\\
    11 & 400  &  0.86 &12  & 800 & 0.85 \\
  \hline
 \end{tabular}
\end{table}

\begin{figure}[hb]
 \begin{subfigure}[b]{\linewidth}
    \includegraphics[width=0.8\hsize]{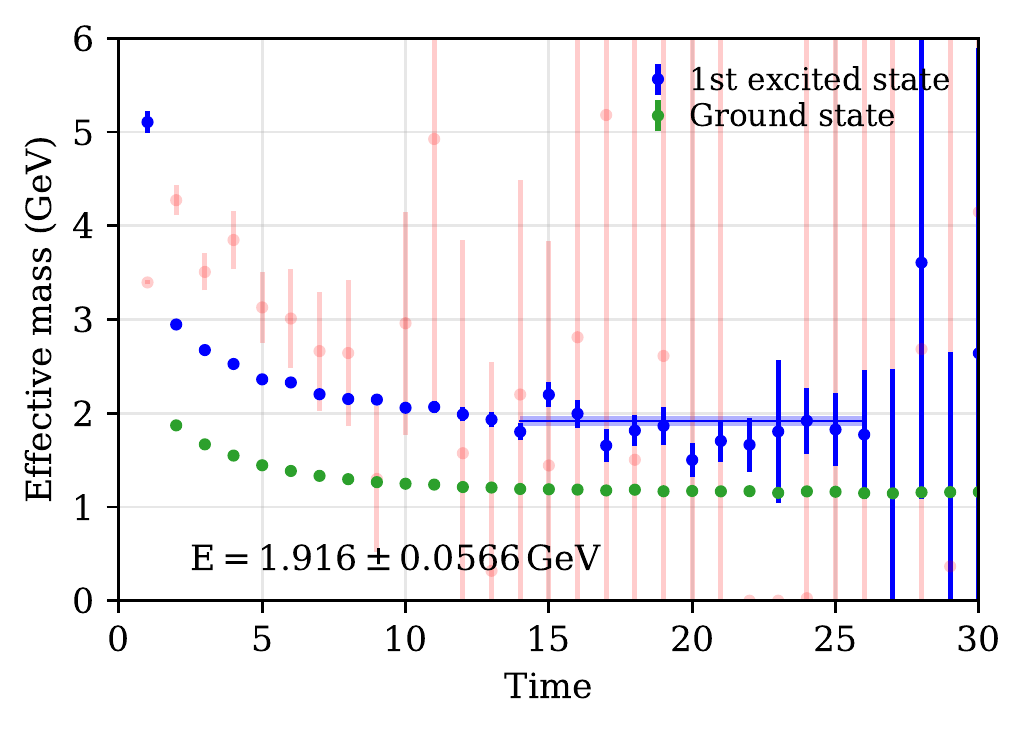}
    \caption{Result from smear parameter $w=0,4,7$, and $11$.}
    \label{fig:var-mass-big}
  \end{subfigure}
  \begin{subfigure}[b]{\linewidth}
    \includegraphics[width=0.8\hsize]{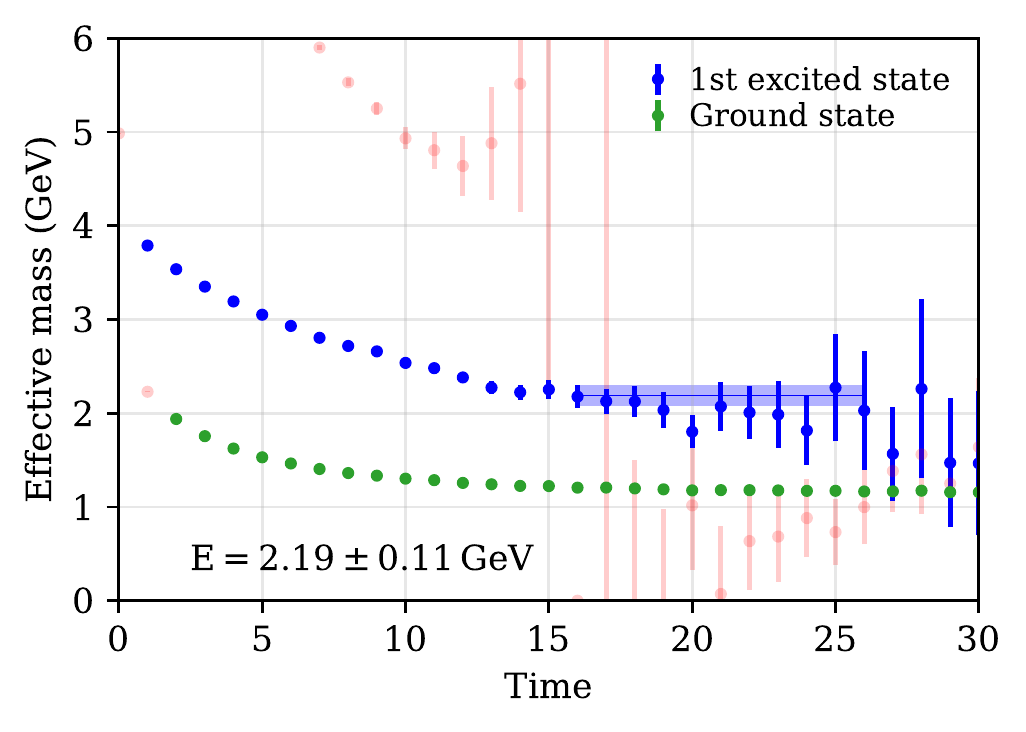}
    \caption{Result from smear parameters $w = 0, 2$, and $4$.}
    \label{fig:var-mass-small}
  \end{subfigure}
  \caption{Effective mass of the states extracted from the variation
    study.  Some states are painted faintly for clarity.  The blue
    horizontal line and band shows the fit of Roper mass.}
  \label{fig:var}
\end{figure}

In view of the fact that the node of the Roper occurs at the radial distance of
$\sim 0.9$ fm for the light pion ($m_{\pi} = 438$ MeV) case, one would want to
have a source with a commensurate size so that it could differentiate the
nucleon from the Roper in the variational calculation. In this regard, we have
chosen 5 operators with different Gaussian smearing sizes whose
root-mean-square (rms) radius ranges from 0 to 0.86 fm, which should be enough
to cover the expected Roper node at $\sim 0.8-0.9$ fm.  Values chosen for the
smearing parameter $w$~\cite{Gong:2013vja} and the number of iterations, and the corresponding rms
radius are listed in Table~\ref{tab:smearing}.  We solve the equation for the generalized
eigenvalue problem (GEVP)~\cite{Michael:1985ne,Luscher:1990ck,Chen:2005mg}
\begin{equation}   \label{gen_eigen}
C(t) v_n (t, t_0)= \lambda(t, t_0) C(t_0) v_n (t, t_0),
\end{equation}
where $C(t)$ is the $N \times N$ correlator matrix for $N$ interpolation field
operators $O_i$ with matrix element defined as
\begin{equation}   \label{gen_eigen_matrix_ele}
C_{ij}(t) = \langle O_i (t) O_j^* (0)\rangle,
\end{equation}
and the $n^{\rm th}$ eigenvalue is expected to be~\cite{Blossier:2009kd}
\begin{equation}
\lambda_n (t, t_0) = e^{-E_n (t - t_0)}(1 + \mathcal{O}(e^{-|\delta E| (t - t_0)}),
\end{equation}
where $\delta E$ is the energy gap between $E_{n+1}$ and $E_n$. We have used
761 configurations of the isotropic clover action, on each of which we chose 36
time slices to place the source, resulting in 27,396 measurements in
total. Spatially, the sources are randomly placed. With the smearing parameters
$w = 0,4,7$ and 11, we find the first nucleon excited state at
$1.92(6)\,$GeV as shown in Fig.~\ref{fig:var-mass-big} for the case $t_0 =
2$. This (red point in Fig.~\ref{fig:roper-dynamical}) is quite consistent with
that from SEB method at $1.87(13)$ GeV (blue point in
Fig.~\ref{fig:roper-dynamical}). On the other hand, when a set of interpolation
fields with only small smearing sizes ($w =0,2$ and 4) are used, we find the
first excited state at $2.19(11)\,$GeV as shown in
Fig.~\ref{fig:var-mass-small} (black point in Fig.~\ref{fig:roper-dynamical}),
which agrees with the JLab results.

This suggests that there are two radially excited states of the nucleon. The
lower one is mainly the 2S state with a radial node at $0.8 - 0.9$ fm as
illustrated in Fig.~\ref{Roper_WF}, while the higher one could be the 3S state
with the first radial node smaller than that of the 2S as evidenced in the
wavefunction study of Ref.~\cite{roberts_nucleon_2014}, which is characteristic
of the Airy function as the solution of a linear potential. As a result, it has
better overlap onto the interpolating source and sink with a smaller size. The
second excited state may well be the $N(1710) 1/2^+$ at 1700 MeV which is $\sim
330$ MeV above the Roper, which is about the same gap we see between the two
excited states in the clover case. Since the variational method with large
smeared operators agrees with the SEB method, it shows that both approaches are
feasible in obtaining the first excited state, namely the Roper.

\begin{figure}[bh]
 \begin{subfigure}[b]{\linewidth}
  \includegraphics[width=0.8\hsize]{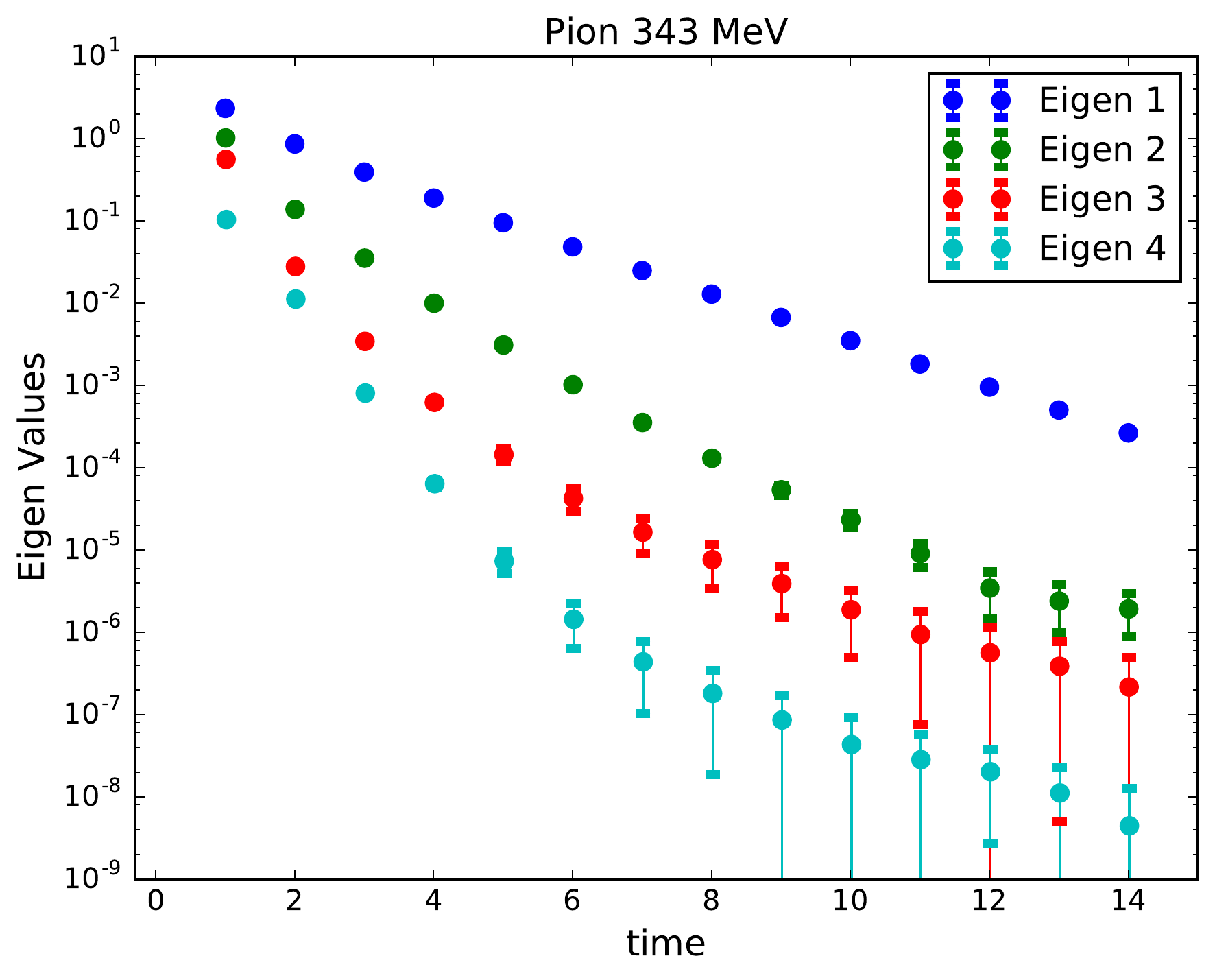}
  \caption{Eigenvalues of the $4 \times 4$ correlation matrix as a function of $t$ for for $m_{\pi} = 343$ MeV.}
    \label{Eigen}  
   \end{subfigure}
  \begin{subfigure}[b]{\linewidth}
  \includegraphics[width=0.8\hsize]{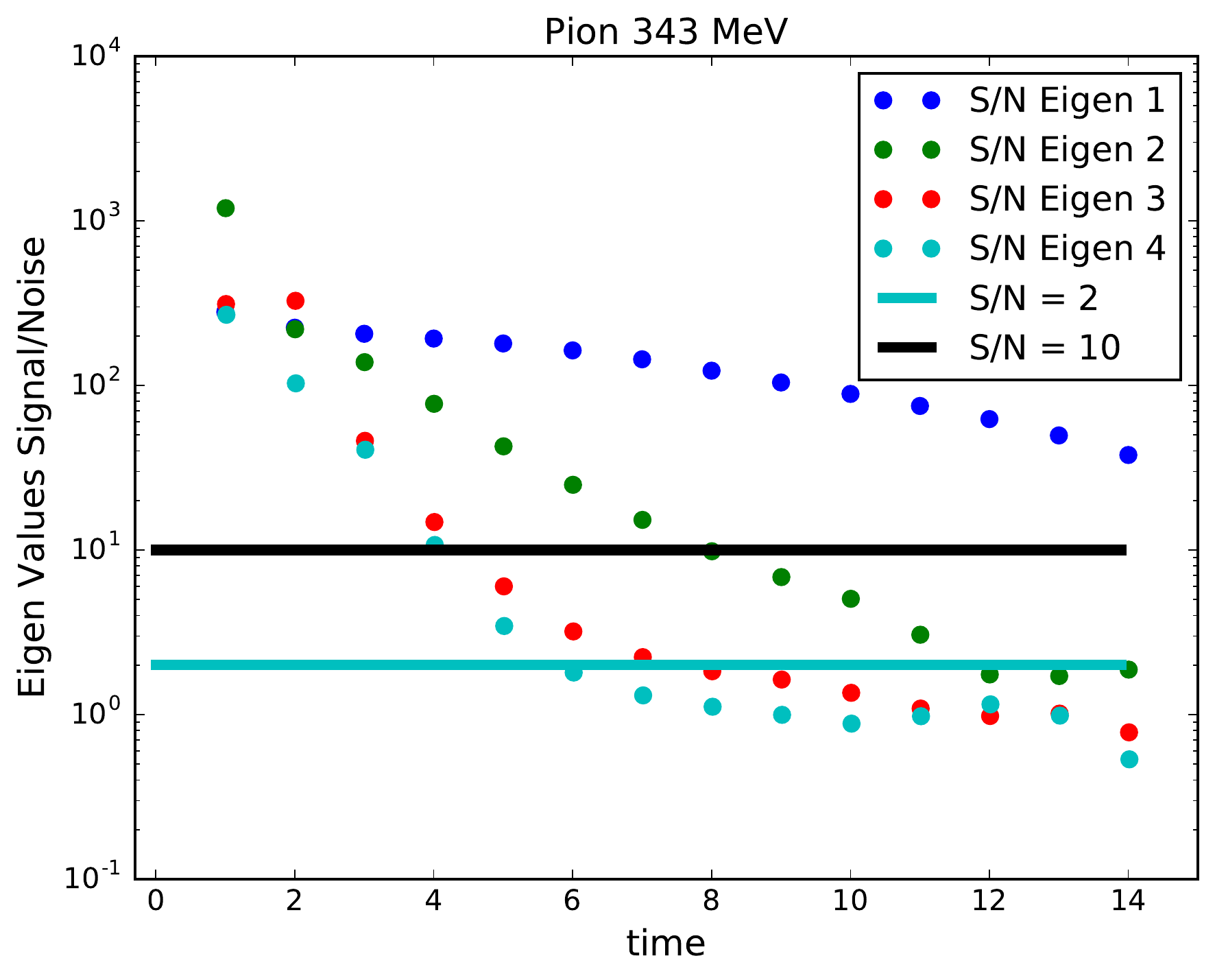}  
  \caption{The signal-to-noise ratio of the 4 eigenvalues as a function of $t$ for $m_{\pi} = 343$ MeV. 
  Also plotted are the S/N = 2 and 10 lines.}
    \label{Eigen_S/N} 
  \end{subfigure}
  \caption{Eigenvalues and the signal-to-noise ratio S/N as a function of time  for $m_{\pi} = 343$ MeV.}
   \label{Eigenvalues}
 \end{figure}

\subsection{III.1 Variational calculation with the overlap fermion}  \label{overlap-variation}

\begin{figure}[ht]
 \begin{subfigure}[b]{\linewidth}
  \includegraphics[width=0.8\hsize]{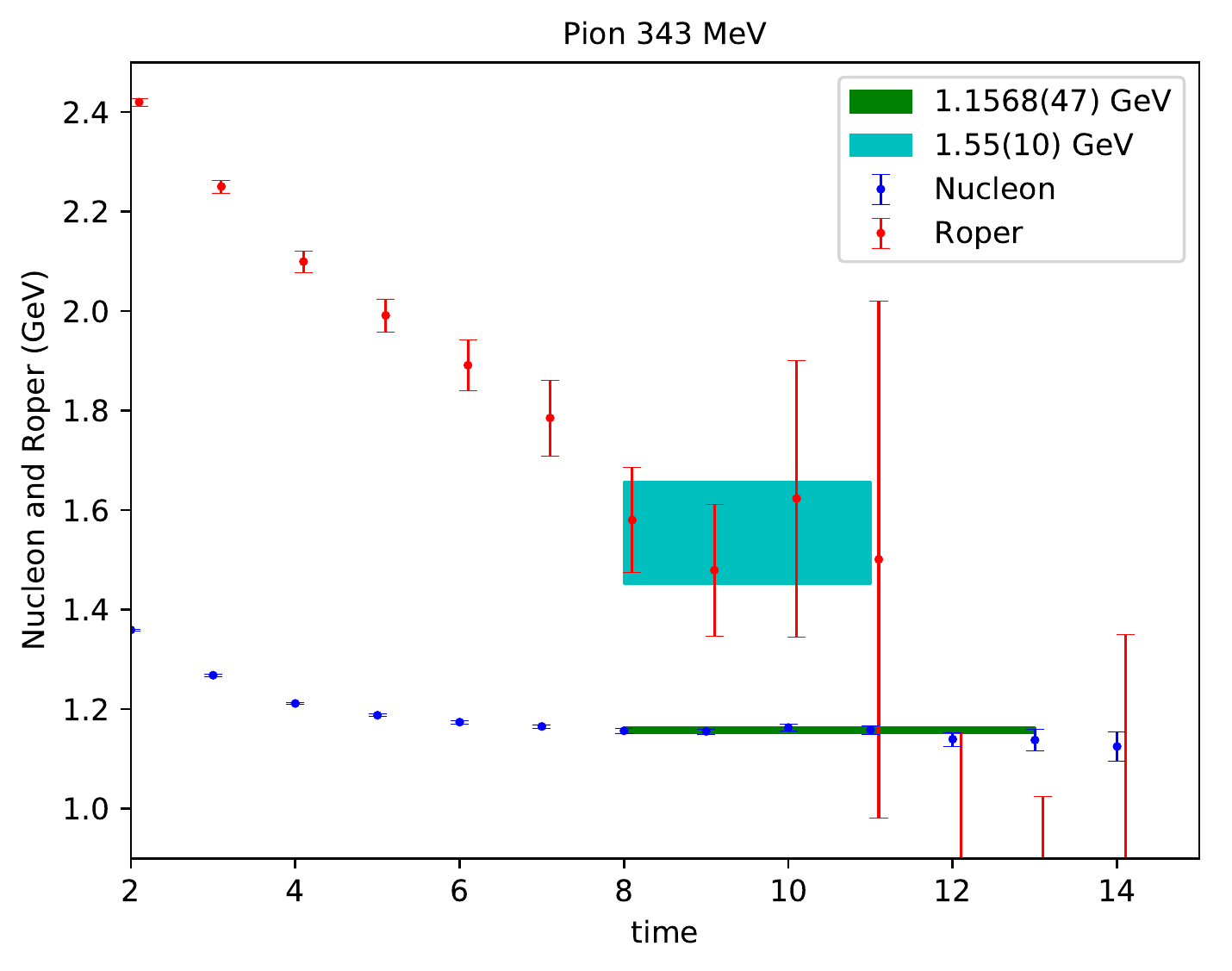}
    \caption{Nucleon and Roper effective masses for $m_{\pi} = 343$ MeV as a function of $t$. The error bands are from the single exponential fits.}
  \label{fix_343}
   \end{subfigure}
  \begin{subfigure}[b]{\linewidth}
  \includegraphics[width=0.8\hsize]{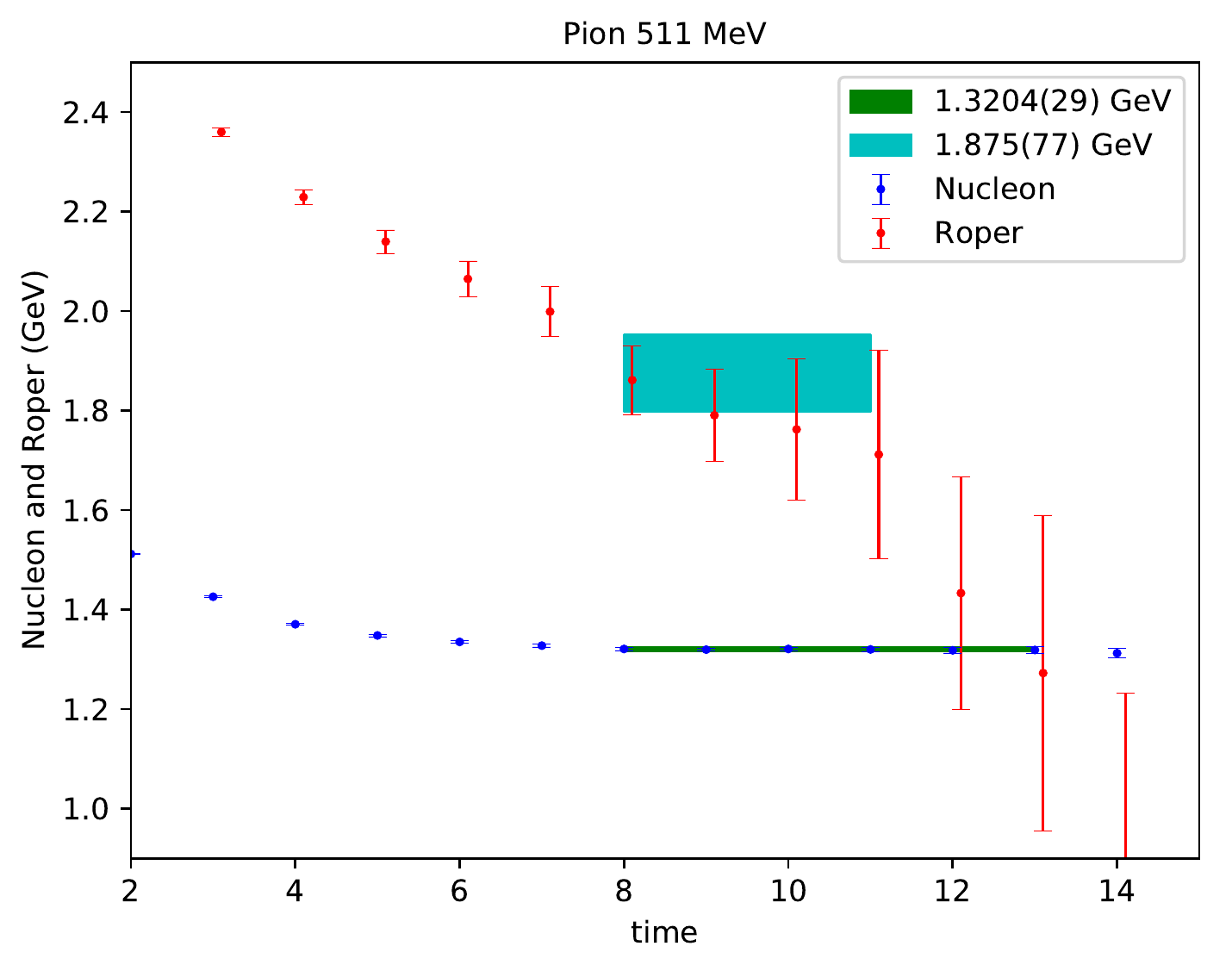}  
   \caption{The same as in Fig.~\ref{fix_343} for $m_{\pi} = 511$ MeV.} 
  \label{fix_511} 
  \end{subfigure}
 \caption{Results of the variational calculations of overlap fermion on $2+1$ fermion DWF configurations for two valence pion masses. These are
 from the eigenvector-projected variation with $w = 0, 2, 4$ and 8 smearing.}
  \label{fig:overlap_fix}
 \end{figure}

Next, we move on to performing variational calculations for the DWF
configurations with valence overlap fermions to check the validity of the
results from the SEB method as alluded to earlier.  We shall first consider the set
of 4 smearing sizes in Table~\ref{tab:smearing} with $w = 0,2,4$ and 8. Before
solving Eq.~(\ref{gen_eigen}), we first determine the combination of
interpolation operators which can lead to well-separated non-degenerate states
as is carried out in Ref.~\cite{Alexandrou:2013fsu}. To do so, we solve for the
eigenvalues and eigenvectors of the $4 \times 4$ correlation matrix $C(t)$ in
Eq.~(\ref{gen_eigen_matrix_ele})
\begin{equation}
C(t) u_n (t)= E_n (t) u_n (t).
\end{equation}
Note that $E_n(t)$ here is not the same as $\lambda(t,t_0)$ in Eg.~(\ref{gen_eigen}).
We plot the eigenvalues of $C(t)$ in Fig.~\ref{Eigen} and their signal-to-noise
(S/N) ratios in Fig.~\ref{Eigen_S/N} as a function of $t$ for the case $m_{\pi}
= 343$ MeV.  We see from Fig.~\ref{Eigen_S/N} that the S/N ratios of the two
lower eigenvalues fall below S/N=2 beyond $t=8$. This is where the first
excited state starts to level off. Thus, we are only able to resolve two states
from this set of smeared operators with the projected matrix
\begin{equation}
\tilde{C}(t) = U^T C(t) U,
\end{equation}
where $U = [u_1, u_2]$ is the $4 \times 2$ matrix spanned by the first two
eigenvectors with acceptable S/N ratios (i.e. $\ge$ 10), for their eigenvalues
at $t = 8$. We then solve the projected GEVP for the projected $2 \times 2$
$\tilde{C}(t)$
\begin{equation}   \label{gen_eigen_proj}
\tilde{C}(t) v_n (t, t_0)= \lambda(t, t_0) \tilde{C}(t_0) v_n (t, t_0),
\end{equation}
at the projection time $t_{\rm{pro}} = 8$ and initial $t_0 =2$. The effective mass is defined as
\begin{equation}
M_n (t) = \ln[(v_n^T(t-1) \tilde{C}(t-1) v_n (t-1))/(v_n^T(t) \tilde{C}(t) v_n (t))]
\end{equation}
where $M_1$ is the nucleon mass and $M_2$ is that of the first excited state
and they are plotted in Fig.~\ref{fix_343} for \mbox{$m_{\pi} = 343$ MeV} and
in Fig.~\ref{fix_511} for the partially quenched valence \mbox{$m_{\pi} = 511$
  MeV.} The bands in these figures show the errors of the fit of $(v_n^T(t)
\tilde{C}(t) v_n (t))$ with a single exponential $e^{-M_n(t - t_0)}$ in the
chosen windows which start at $t = 8$. It is gratifying to see
that the Roper mass $M_R = 1.55 (10)$ GeV at \mbox{$m_{\pi} = 343$ MeV} is in
good agreement with the corresponding value of 1.60(12) GeV from the SEB
method.

\begin{table}[ht]  
\caption{Test of sensitivity to the size of the Gaussian source. A Coulomb and a Gaussian source with three sizes are tested. Gauge-invariant box
sinks $B_N$ (N = 2, 4, 6, 8, 12 ) are listed for each of the three combined Coulomb and smeared sources. 
}
 \label{tab:9_methods}
\begin{tabular}{| c | c | c | c | c | c |}
\hline
Method &  Source      & Sink                & $t_0$   & $t_{\rm ref}$  \\
\hline
1     &  Wall, $\omega$12    & B2,B6           & 5  &  6     \\
2      &  Wall, $\omega$12    & B4,B8         & 5  &  6     \\
3      &  Wall, $\omega$12    & B8,B12         & 4  & 6     \\
4     &  Wall, $\omega$8     & B2,B6         & 4  &  6     \\
5      &  Wall, $\omega$8     & B4,B8         & 4  &  6     \\
6      &  Wall, $\omega$8     & B8,B12         & 4  & 6     \\
7      &  Wall, $\omega$5.5   & B2,B6          & 3  &  6     \\
8      &  Wall, $\omega$5.5   & B4,B8         & 3  &  6     \\
9     &  Wall, $\omega$5.5   & B8,B12         & 3  &  6     \\
\hline
\end{tabular}
\end{table}

\begin{figure}[ht]
 \begin{subfigure}[b]{\linewidth}
  \includegraphics[width=0.8\hsize]{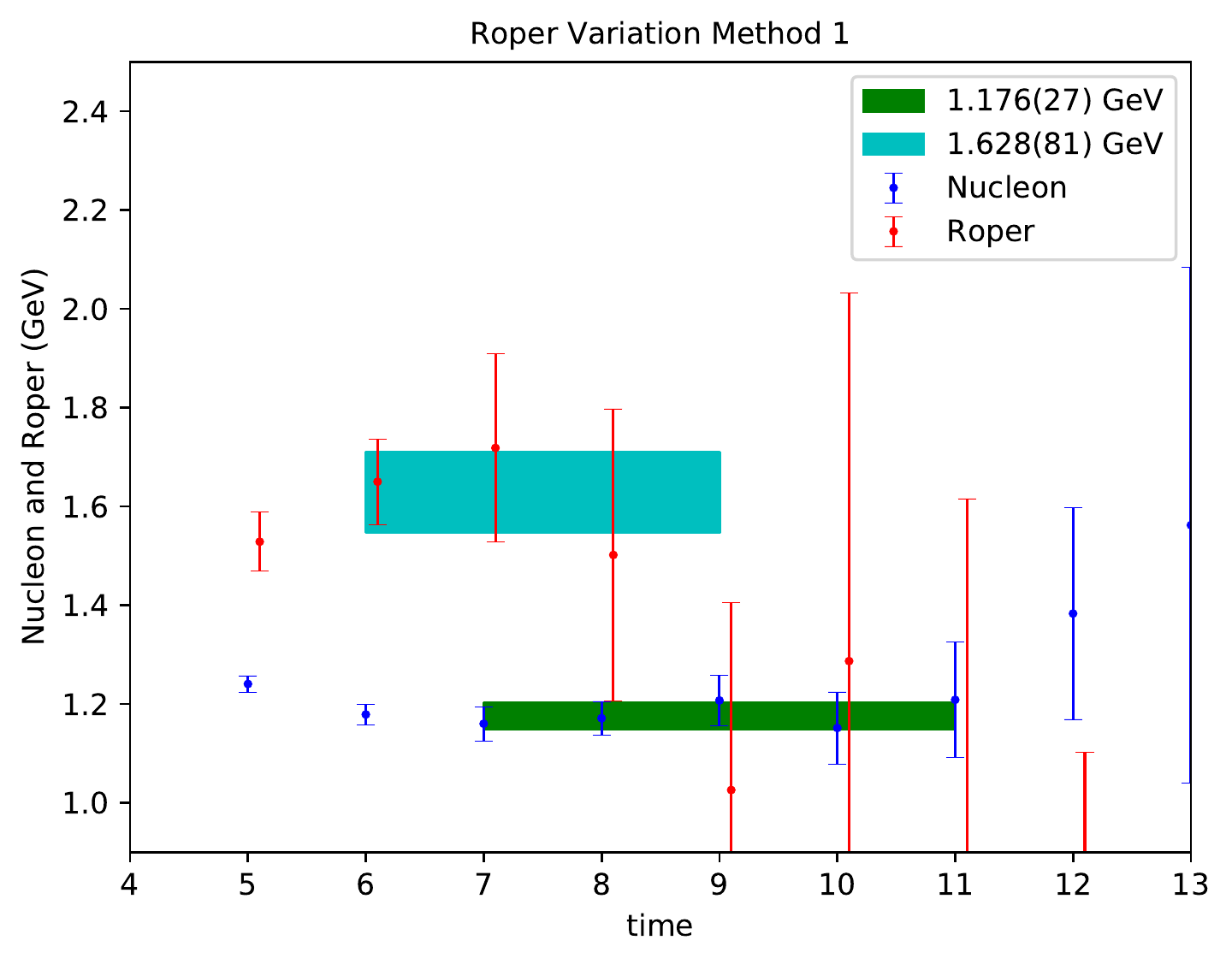}
    \caption{Nucleon and Roper effective mass for method 1 in Table~\ref{tab:9_methods}.}
  \label{Wall_M1}
   \end{subfigure}
  \begin{subfigure}[b]{\linewidth}
  \includegraphics[width=0.8\hsize]{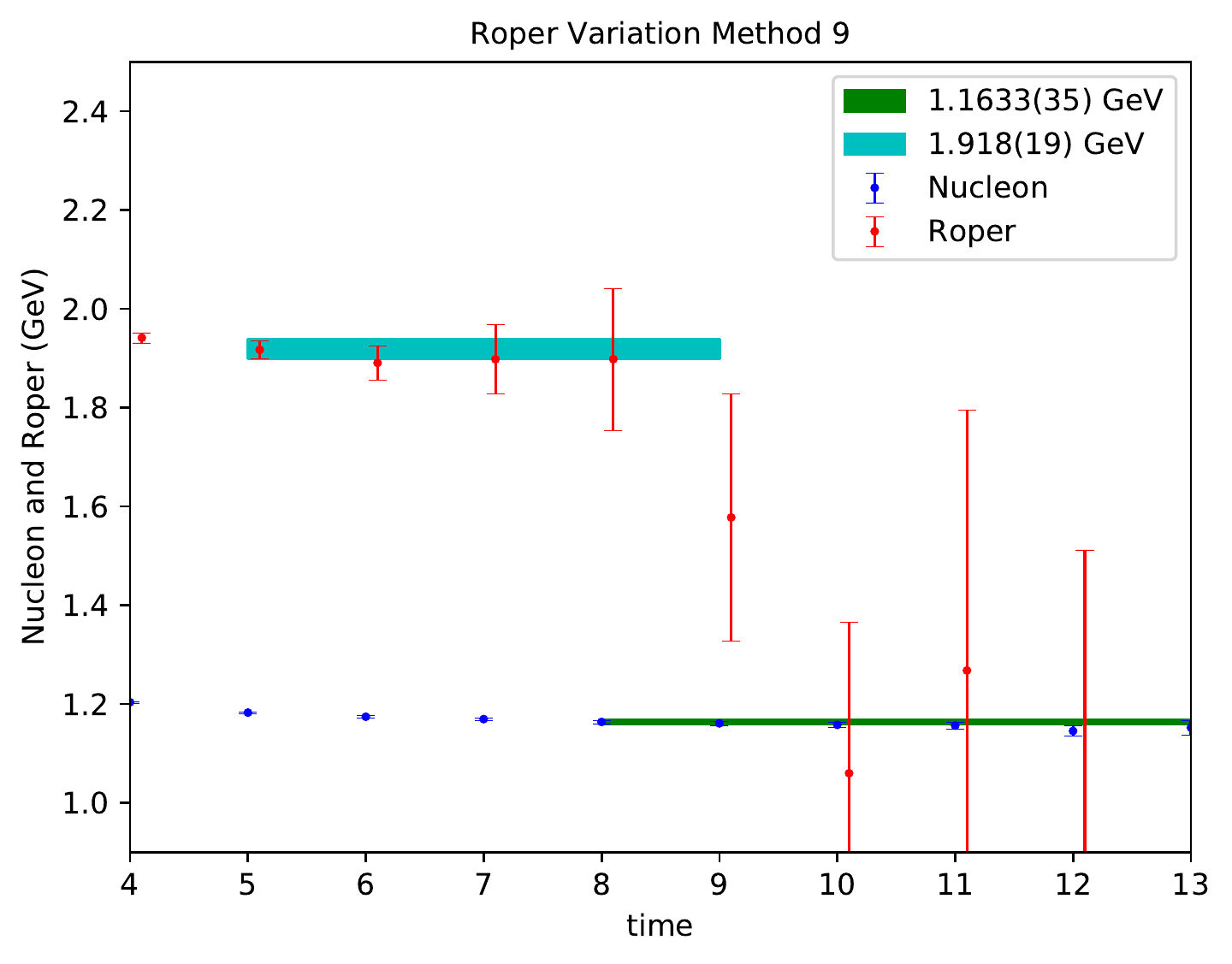}  
   \caption{Same as in Fig.~\ref{Wall_M1} for method 9.} 
  \label{Wall_M9} 
  \end{subfigure}
   \caption{Results from variational calculation of overlap fermion on $2+1$ fermion DWF configurations at $m_{\pi} = 343$ MeV with
   different sources and sinks whose specifics are given in Table~\ref{tab:9_methods}.}
   \label{fig:overlap_wall}
 \end{figure}

\begin{figure}
    \includegraphics[width=0.8\linewidth]{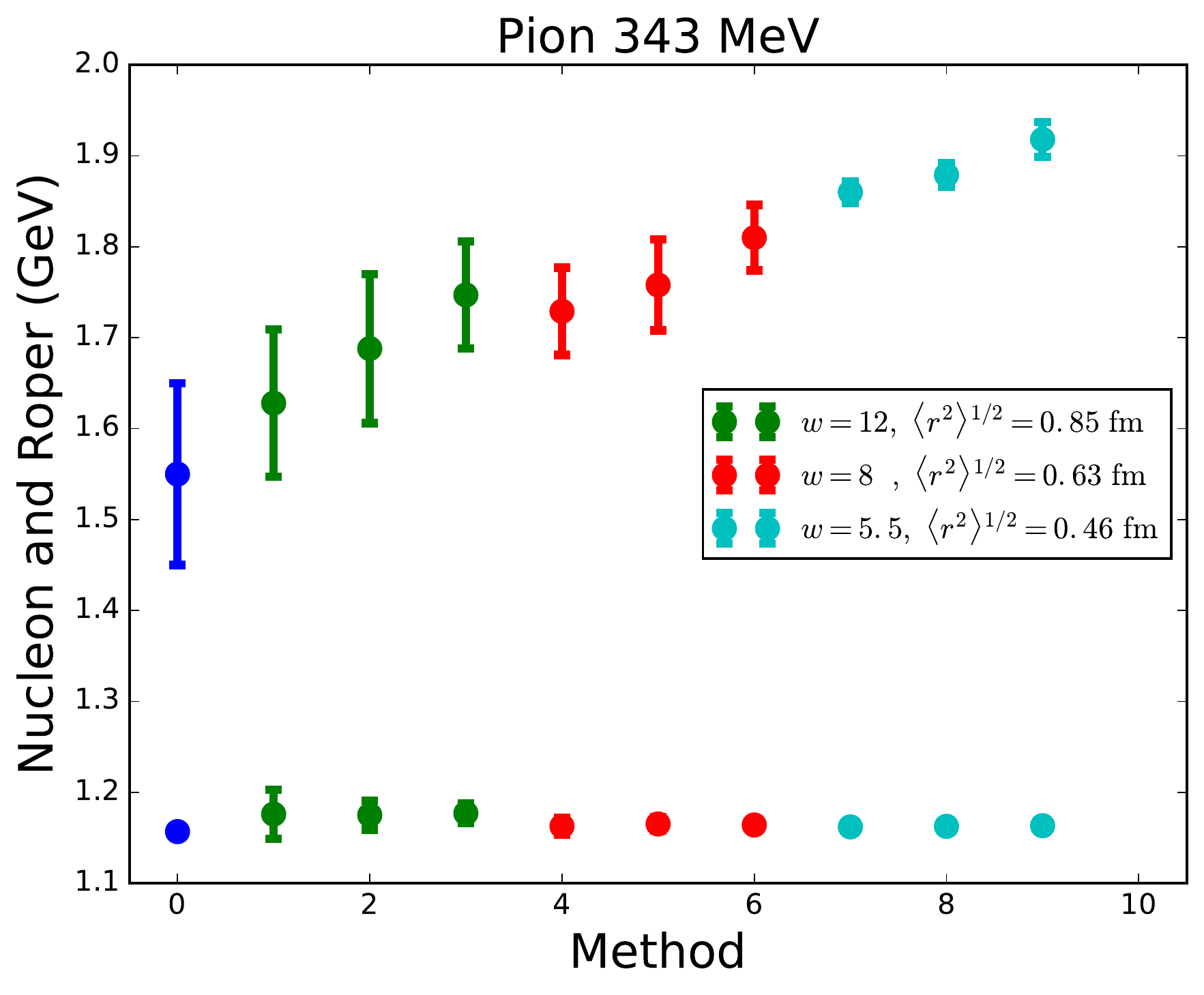}
    \caption{The nucleon and first excited state in the $1/2^+$ channel for 9 methods with various sources and sinks listed in Table~\ref{tab:9_methods}.
    Method 0 is the case with eigenvector-projected variation with $w = 0, 2, 4$ and 8 smearing. These are for the case $m_{\pi} = 343$ MeV.}
  \label{fig:methods}
\end{figure}

To check the sensitivity of the excited state to the smearing size and the
related issue of a possible second excited state, as alluded to in the study of
the anisotropic clover lattice discussed in Sec.~III, we employ additional
variational calculations with different smearing sizes for the nucleon
source. We consider a combination of a Coulomb wall and a Gaussian smeared
source and gauge-invariant box smeared sinks~\cite{Liang:2016fgy} in the
variational calculation. The gauge-invariant box smearing is more economical
than Gaussian smearing, especially when the smearing size is
large~\cite{Liang:2016fgy}. We shall use multiple box sizes for the sink. This entails variational
calculations with asymmetric correlation matrices. Variational approach with asymmetric correlation
matrices has been applied to the excitation spectrum of baryons, particularly the
positive- and negative-parity excited states of the nucleon~\cite{Melnitchouk:2002eg,Mahbub:2009nr,Mahbub:2010me}.
The size of the smeared box $B_N$ is $N$ lattice spacings and the physical length is $L_{B_N} = N a$, where
\mbox{$a = 0.1105$ fm} for this lattice. Besides the wall source, we have
another Gaussian smeared source with $w =5.5, 8$ or 12. The rms radii for the
sources with $w =8$ and 12 are 0.63 fm and 0.85 fm, respectively, as listed in
Table~\ref{tab:smearing}, and that of $w = 5.5$ is 0.46 fm. In each case, we
used two box sources ($B_N$) to solve the generalized eigenvalue problem (GEVP) for the $2 \times 2$ asymmetric correlation matrix 
for each $t > t_0$. The specifics of the source and sink, and $t_0, t_{ref}$ are given for a total
of 9 different setups in Table~\ref{tab:9_methods}. The eigenvalues $e^{-m_it}$ and eigenvectors are real for the 
asymmetric correlation matrix so defined. The effective mass and $(v_n^T(t) \tilde{C}(t) v_n
(t))$ are defined from the eigenvectors at $t_{\rm ref}$ as specified in Table~\ref{tab:9_methods}.
It turns out that all the methods with different source sizes produce the same
nucleon mass for $m_{\pi} = 343$ MeV, but a range of Roper masses from 1.63(8)
GeV for method 1 to 1.92(2) GeV for method 9. The effective masses for the
nucleon and Roper for methods 1 and 9 are plotted in
Fig.~\ref{fig:overlap_wall}, similar to those in Fig.~\ref{fig:overlap_fix}. We
present the nucleon and the Roper results in Fig.~\ref{fig:methods} for the 9
methods in Table~\ref{tab:9_methods} together with that with 4 sources and 4
sinks from Fig.~\ref{fix_343}. We see that, again, the sources with large
smearing ($w=12$ for method 1, 2 and 3) (colored green) have lower masses than
those with smaller smearing (red points with $w = 8$ ) and still smaller
smearing (cyan points with $w = 5.5$). This is basically the same as we found
with the anisotropic clover lattice in Sec.~III, i.e. the use of the smaller
smearing sources results in a higher value for the measured excited-state
mass. This is consistent with the picture that there are two radially excited
states, namely the Roper and $N(1710) 1/2^+$ at 1700 MeV, which is $\sim 330$
MeV above the Roper.  The latter, being 3S radial excitation with a smaller
node position overlaps better with a source of a smaller size. In
Fig.~\ref{fig:methods}, an almost monotonic increase of the measured value of
the excited-state mass as the source size decreases. This suggests that the
small $2 \times 2$ correlation matrix results in a contamination between the
first and second excited states in the windows amenable to fits with reasonable
errors. The clear signal in Fig.~\ref{Wall_M9} suggests that it is the pure
second excited state. With the present setup and statistics, we are not able to
resolve 3 separated states with the chosen set of operators. Otherwise, it
would have been easier to verify the picture that we infer.

We add the results of the variational calculation with $w = 0, 2, 4$ or 8 on
Fig.~\ref{fig:roper-dynamical} (brown points) for several pion masses to show
that they are consistent with those from the SEB method. The upshot of the
variational calculation with large smearing size is that its results are
consistent with those from the SEB method for both the clover and overlap
fermions. This removes the uncertainty regarding the fitting algorithms. The
remaining challenge is to understand why the Roper with overlap fermion is
lower than that from the clover fermion by $\sim 300$ MeV on lattices with
similar lattice spacing and sea quark mass.

{We have not studied the source-size dependence of SEB, but we presume the size will have an effect, similar to what is found in the variational approach. In this work, we used the Coulomb wall source for the overlap data on the $24^3 \times 64$ lattice in Sec. II which are shown in Fig. 2 as the pink points. The wall source is known to suppress the higher excited states so that the first excited state can be extracted more readily in the fitting algorithm, such as SEB. Had we used a point source, the SEB would have given a high first excited state, similar to that of the variational approach with small smearing sources. This is so because all the nS radially excited states would have contributed coherently to the spectral weights for the point source and the present lattice spacing at $0.114$ fm is not fine enough to isolate the 2S state from the 3S and higher excited states. It is conceivable that varying the size of the source in SEB may well result in a mixture of 2S and 3S states as shown in the variational approach in Fig. 8. We will study this in the future.}

{We should mention that the Roper is a resonance which decays to $N\pi, N\pi\pi, \Delta\pi \cdots$... To find its pole mass and width on the lattice, one should do the scattering phase shift calculation including the $N\pi$ channel with the Luscher method, such as those that have been done in Refs.~\cite{Lang:2016hnn,Kiratidis:2017bnr}. The Luscher method relates the interaction energy shift of the two hadrons to the phase shift. In the case of the Roper, this energy shift is a reflection of the mixing between the confined would-be Roper from the 3-quark interpolation field and the $\pi N$ P-wave scattering state at a particular relative momentum. As the two states approach each other when the relative momentum between the $\pi$ and $N$ changes, there is level repulsion to avoid crossing. The maximum repulsion occurs at maximum mixing of the two states and they are positioned at the half maxima of the scattering cross section (or phase shift $\delta$ = 90 degrees) sandwiching the maximum point for a Breit-Wigner resonance cross section, which reflects the full-width-at-half-maximum (FWHM). When the width of the resonance is relatively narrow (e.g. the Roper width is $\sim$ 175 MeV) compared to its mass, the maximum repulsion occurs close to the would-be crossing point between the non-interacting two hadron state (i.e.\ $\pi N$) and the confined would-be Roper from the 3-quark interpolation field. When the lattice volume is large enough so that the would-be Roper does not change appreciably with increasing volume, the would-be crossing point is just the mass of the would-be Roper. In this case, the Roper mass from the 3-quark interpolation field in this work should be close to the resonance pole calculated from the $\pi N$ scattering study \`a la Luscher. The shift of the pole position from the would-be Roper mass is related to the deviation from the Breit-Wigner form and is bounded by half of the width $\Gamma/2$ which is $\sim 88$ MeV for the Roper. This is much smaller than the $\sim 600$ MeV discrepancy between the Roper states calculated from the clover and overlap fermions, which is what we try to resolve in this study. We can find evidence for the above argument from the lattice calculations of the $\rho$ resonance from $\pi\pi$ scattering. One is a $N_f =2$ calculation of the $\rho$ resonance from $\pi\pi$ scattering in the elongated lattice to change the pion momentum~\cite{Guo:2016zos}. The resonance mass $a m_{res} = 0.4878(4)$ turns out to be very close to the would-be $\rho$ mass of $a m_{\rho} = 0.4800(30)$~\cite{Brett:2020} from the $\bar{q}\gamma_iq$ interpolation field at a volume where the $\pi\pi$ P-wave state is far above the would-be $\rho$. The other $\pi\pi + K\bar{K}$ calculation ~\cite{Wilson:2015dqa} is carried out on an isotropic lattice for the $N_f = 2+1$ case; the resonance mass is again very close to that from the $\bar{q}\Gamma q$ interpolators. Since the width of the Roper at 175 MeV is close to that of the $\rho$ at 150 MeV, we think that using 3-quark interpolation field to calculate the Roper mass in this work should be a good approximation to that of the Roper resonance.}

\section{IV. Multi-hadron state from single-hadron interpolation field}

To track down the origin of the difference in the prediction of the Roper mass
in clover and overlap fermions, we note that chiral effective theory hints
toward substantial mixtures of higher Fock-space components, involving $N \pi$, 
and $N \pi\pi$ states, as the reason why the quark model's Roper prediction
is too high. As mentioned in the introduction in Sec. I, the most sophisticated
dynamical coupled-channel (DCC)
model~\cite{JuliaDiaz:2007kz,Kamano:2010ud,Suzuki:2009nj} shows that the
coupling to $\pi N$, $\eta N$ and $\pi\pi N$ brings down the uncoupled bare
Roper state by $\sim 400$ MeV to the experimental value. The experimental
electroexcitation amplitudes of the Roper resonance provides evidence for it to
be primarily the radial excitation of a three-quark core, while higher Fock
space components are needed to describe the low $Q^2$ behavior of the
amplitudes~\cite{Aznauryan:2011qj}.
   
\begin{figure}
    \includegraphics[width=0.6\linewidth]{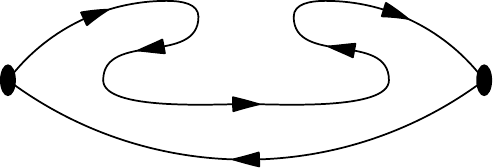}
    \caption{Hairpin diagram in the quenched isovector scalar meson ($a_0$)
      correlator. When the pion mass is lower than $\sim 600$ MeV, it produces
      a ghost would-be-$\eta\pi$ state which shows up as a negative tail at
      large time separation for the overlap fermion.}
  \label{fig:hairpin}
\end{figure}

To study the implication of the higher Fock space and the difference
between the clover and overlap fermions, we note that there is a
well documented would-be $\eta\pi$ ghost state in the isovector scalar meson
channel (i.e., the $a_0$ channel) in the quenched approximation with overlap
fermion~\cite{mathur_lattice_2007}. This is caused by the ``hairpin'' diagram
as illustrated in Fig.~\ref{fig:hairpin} with the $\bar{\psi}\psi$
interpolation field.

\begin{figure} [htb]
    \includegraphics[width=1.0\linewidth]{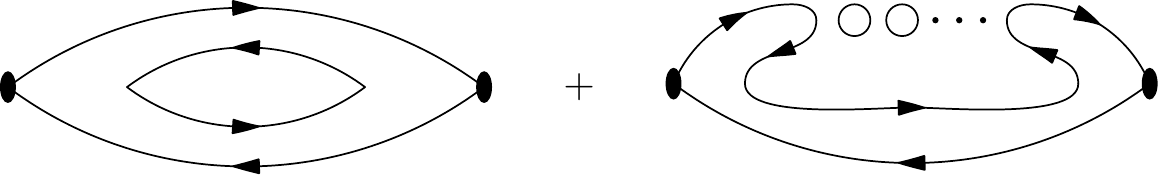}
    \caption{The isovector scalar meson ($a_0$) correlator with dynamical
      fermion configurations. The quark loops in the vacuum are responsible for
      producing full $\eta\pi$ and $\eta^\prime\pi$ propagators in the
      $2+1$-flavor case.}
  \label{fig:eta-pi}
\end{figure}

It is clearly observed that the $a_0$ correlator starts to develop a negative
tail when the pion mass is less than $\sim 600$ MeV, and it is progressively
more negative at earlier time slices for smaller quark masses. This is a clear
indication that at least one of the ghost $\eta\pi$ states, being lightest in
mass, is dominating the correlator over the physical $a_0$ at larger
times. This has been reported in the literature for the
quenched~\cite{Bardeen:2001jm,Bardeen:2003qz,Prelovsek:2002qs} and partially
quenched~\cite{Prelovsek:2004jp} calculations.  This ghost $\eta\pi$
contribution in the $a_0$ correlator has been studied in the chiral
perturbation
theory~\cite{Bardeen:2001jm,Prelovsek:2004jp,mathur_lattice_2007}. The Fourier
transform of the one-loop hairpin diagram gives the following contribution to
the isovector scalar meson propagator in Euclidean time~\cite{mathur_lattice_2007}
\begin{eqnarray}  \label{G-S}
&&G_S(\vec{p}=0) =  FT \{a^2\Delta_h(\vec{p}=0)\}  \nonumber \\
&=& - \frac{r_0^2m_0^2}{2N_S^3}
 \frac{(1+m_{\pi}t)}{2m_{\pi}^4} e^{-2m_{\pi}t} + (t \rightarrow N_T -t),
\end{eqnarray}
where the $(1+m_{\pi}t)$ factor is due to the double pole of the would-be
$\eta$ ghost propagator in the loop.  $r_0$ is the coupling between the scalar
interpolation field and the $\eta$ and $\pi$ or the matrix element $\langle
0|\bar{\psi}\psi|\eta\pi\rangle$ and $m_0^2 = 2 N_f \, \chi_t/f_{\pi}^2$
is the hairpin insertion mass which is related to the
topological susceptibility $\chi_t$ in the pure gauge theory.  Note this ghost
state contribution is negative due to the coupling between two annihilating
pseudoscalar propagators in the unfinished $\eta$ channel without the presence of
quark loops as in Fig.~\ref{fig:eta-pi}.

In a full dynamical fermion situation, the quark loops from the fermion
determinant give rise to a geometrical series to lift the would-be Goldstone
boson to $\eta$ and $\eta^\prime$ in the Witten-Veneziano formalism to resolve
the $U(1)$ anomaly~\cite{Witten:1979vv,Veneziano:1979ec}.  This is illustrated in
the cartoon picture in Fig.~\ref{fig:eta-pi}. In the quenched approximation,
there are no quark loops. This results in the hairpin diagram from the double
annihilation of the would-be $\eta$ in Fig.~\ref{fig:hairpin} and gives the
negative ghost $\eta\pi$ behavior in the quenched $a_0$
correlator~\cite{Bardeen:2001jm,Prelovsek:2004jp,mathur_lattice_2007}. The
existence of this ghost propagator is a prelude to revealing the fact that the
$\eta\pi$ type multi-hadron states can be produced by the one-hadron $\bar{q}q$
interpolation field in a full dynamical fermion calculation.  Even though this
ghost $\eta\pi$ behavior is well-documented in a quenched calculation with the
overlap fermion for a range of pion mass from 600 MeV down to 180 MeV, there
has not been such clear evidence for the Wilson-type fermion. We shall make a
comparison between the overlap fermion and the clover fermion.

\begin{figure} [ht]
\begin{subfigure}[b]{\linewidth}    
  \includegraphics[width=0.8\hsize]{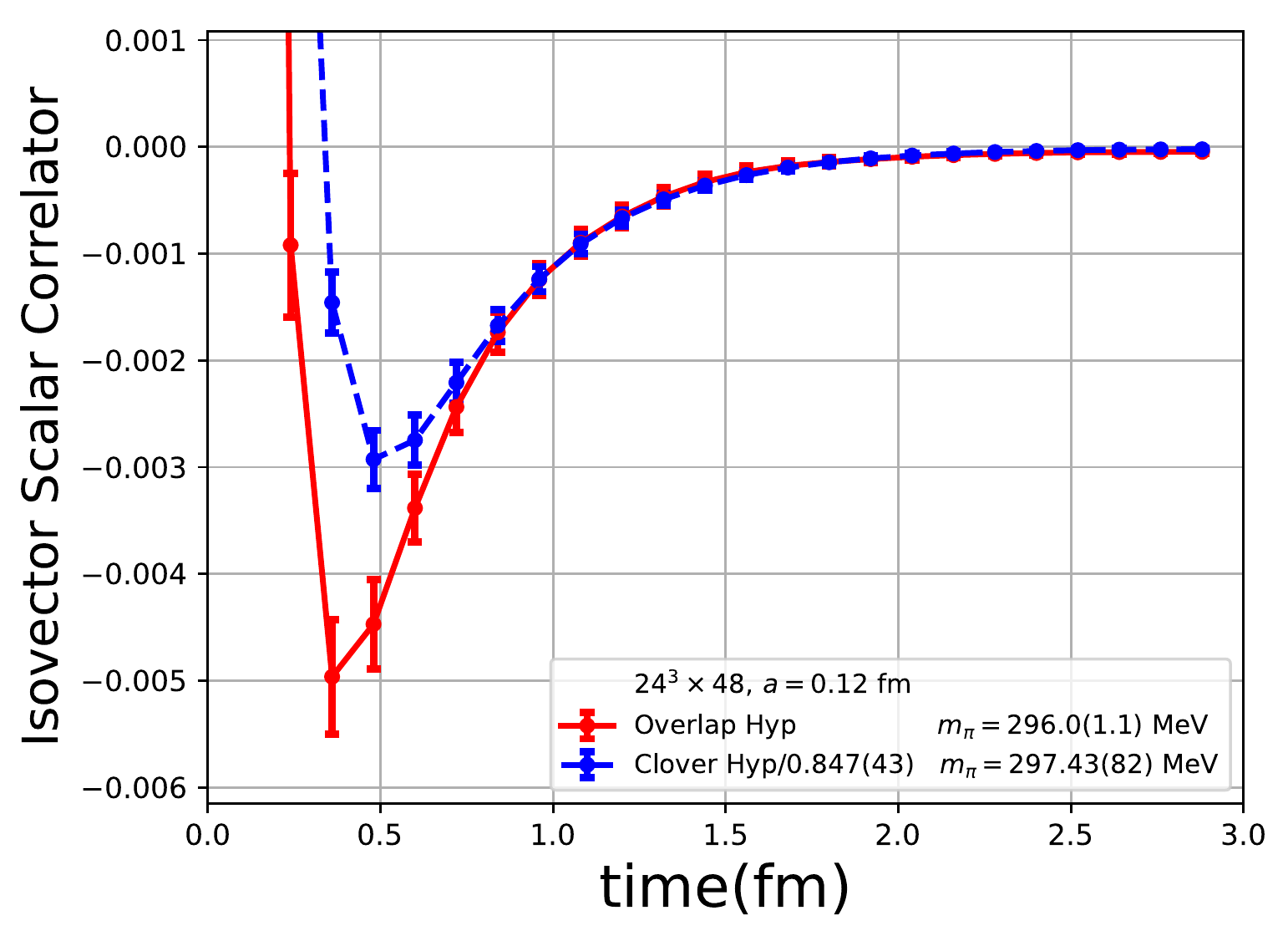}
\end{subfigure}
\begin{subfigure}[b]{\linewidth}
       {\includegraphics[width=0.8\hsize]{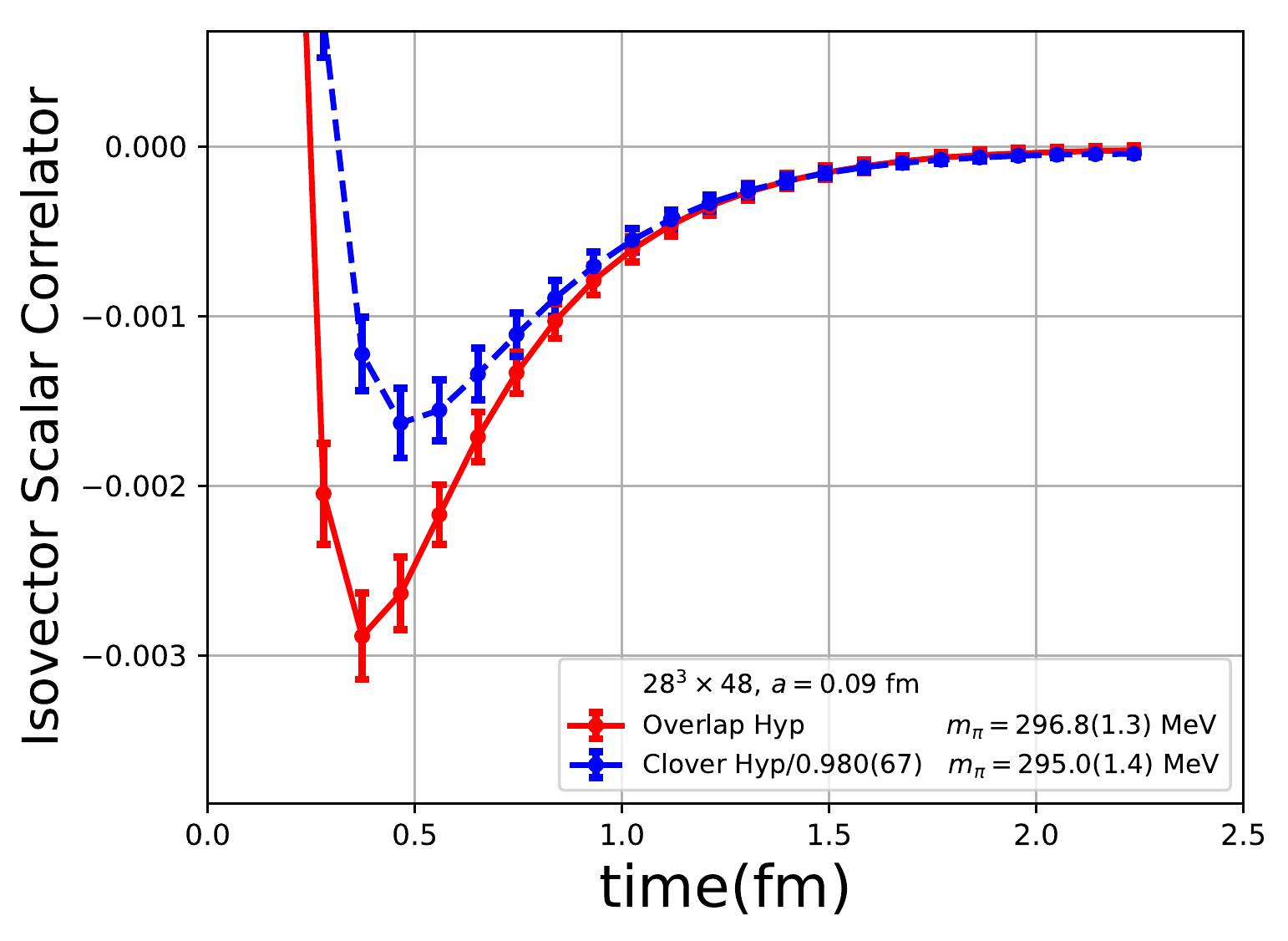}}
\end{subfigure}
\begin{subfigure}[b]{\linewidth}
 {\includegraphics[width=0.8\hsize]{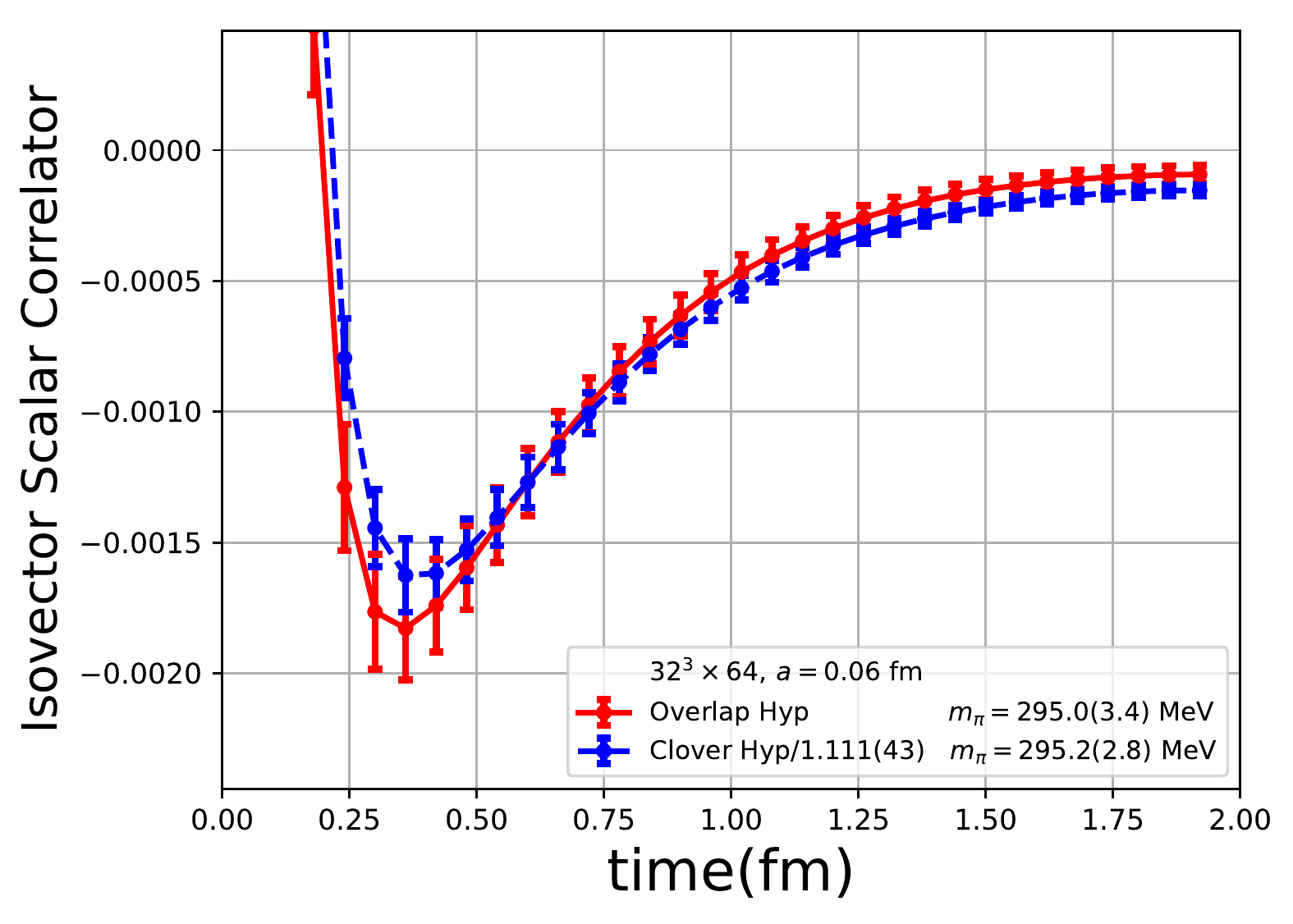}}
\end{subfigure} 
     \caption{Comparing the isovector scalar meson ($a_0$) correlators between clover and overlap fermions for pion masses at $\sim 296$ MeV for lattice spacings $a = 0.12$ fm (upper panel), $ a = 0.09$ fm (middle panel) and
$a = 0.06$ fm (lower panel). The lines connecting the points are there to guide the eyes.}
  \label{fig:ghost_clo/ov}
\end{figure}

Plotted in Fig.~\ref{fig:ghost_clo/ov} are the 
$a_0$ correlators on three quenched Wilson-gauge lattices with lattice size/spacing of 
$24^3 \times 48/0.12\, {\rm fm}$ (upper panel), $28^3 \times 48/0.09\, {\rm fm}$ (middle panel), 
and $32^3 \times 64/0.06\, {\rm fm}$ (lower panel),
respectively. The $a_0$ correlators are calculated at $\sim 296$ MeV pion mass
for the overlap and clover fermions with the mean-field clover term. Both of them are HYP-smeared. 
To take into account the fact that the
renormalization factors $Z_S$ are different in the two actions, we compare
their $a_0$ correlators for the heavy pion case ($\sim 1$ GeV) where there are no ghost
states. To normalize the correlators for comparison in Fig.~\ref{fig:ghost}, we divide the
clover $a_0$ correlator by a factor which is the ratio of the clover to overlap
correlator values at large time separation with the $a_0$ mass at $\sim 1.6 - 1.9$ GeV.
The factors are $0.847(43), 0.980(67), 1.111(43)$ for the lattices with spacings $0.12,
0.09$ and 0.06 fm, respectively. 

The $\eta \pi$ ghost states are clearly seen in all three lattices where the correlators turn negative beyond
$t \sim 0.25$ fm.The conspicuous feature of the comparison is that the negative parts of the clover correlators for the cases of $a = 0.12 $ fm (upper panel) and $a = 0.09$ fm (middle panel) are shallower than those of the overlap,
whereas they are about the same at $a = 0.06$ fm (lower panel).
The existence of the minimum is the result of a positive exponential term from the physical $a_0(1236)$ and a negative
exponential term from the $\eta \pi$ ghost in Eq.~(\ref{G-S}). 
Since we have taken the relative renormalization of the clover and the overlap into account from comparison of
the $a_0$ correlator at large pion mass where there is no ghost state, the physical $a_0$ exponential should be
the same for the cases in Fig.~\ref{fig:ghost_clo/ov}. Their differences should reflect the couplings to the $\eta \pi$ ghost state. Therefore, we take the correlator at the minimum to serve as an approximate yet meaningful indicator for the strength of coupling to the $\eta \pi$ state. 
\begin{figure} [ht]
      \includegraphics[width=0.7\linewidth]{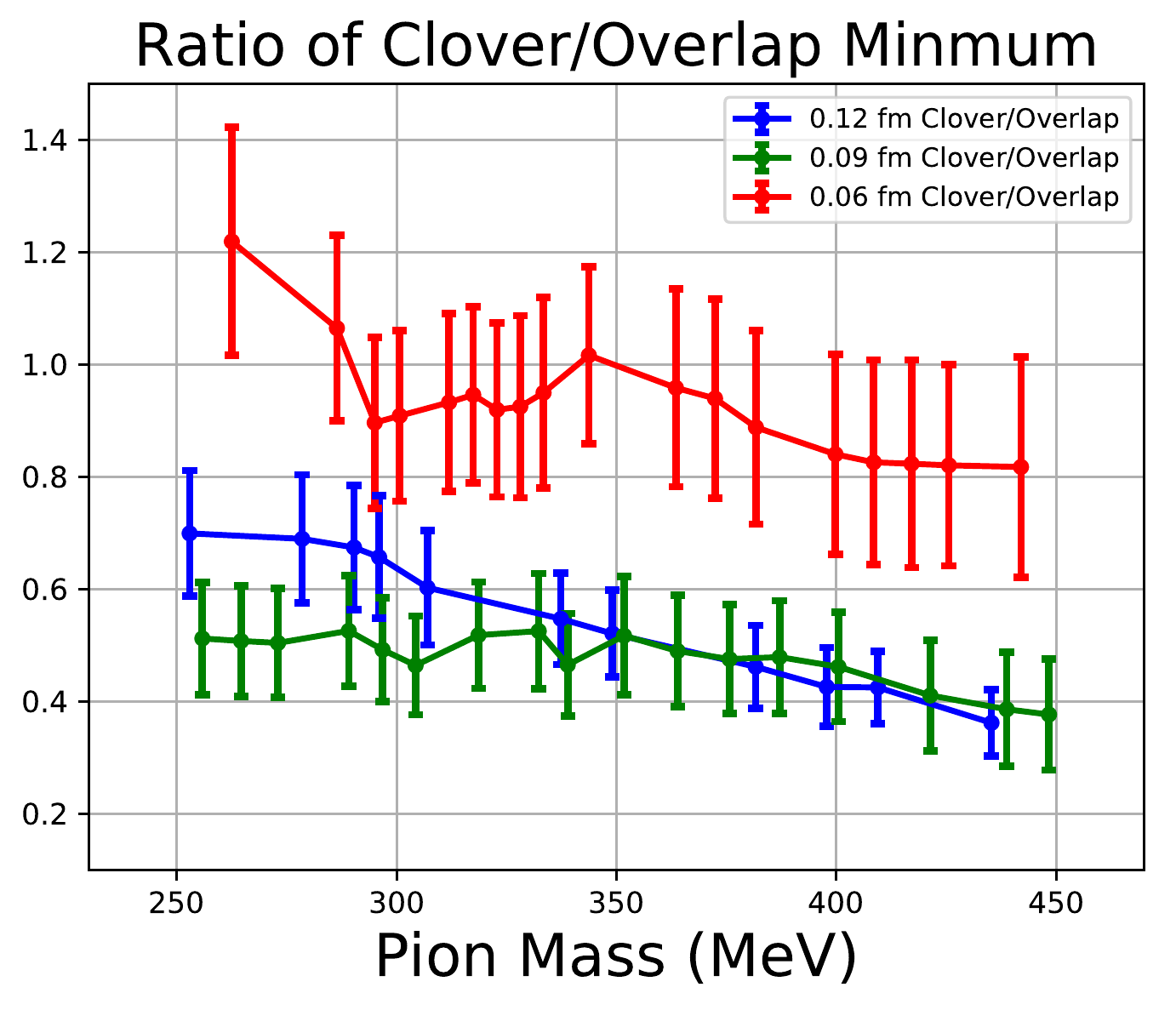}
    \caption{The ratios of the minima of the quenched $a_0$ correlator of the clover fermion to that of overlap fermion for a range of pion masses in three lattice spacings. }
  \label{fig:min_compare_clover}
\end{figure}
To see the quark mass dependence, we calculate the ratio of the $a_0$ correlator between the clover and overlap fermions at their respective minimum in the pion mass range between 250 MeV and 450 MeV and plot in
Fig.~\ref{fig:min_compare_clover}. We see that for $a = 0.12$ fm and 0.09 fm the ratios are about
0.4 $-$ 0.6 and the pion mass dependence in this range is not strong. On the other hand, the ratio is close to unity 
for the $a = 0.06$ fm case across the same mass range. This basically reflects what we have seen in Fig.~\ref{fig:ghost_clo/ov}.
 It is commonly believed that chiral symmetry is recovered for the Wilson-type fermion in
the continuum limit; we interpret our results to imply that the chiral symmetry breaking effect of the clover fermion
is indeed observable at $a = 0.12$ fm and $a = 0.09$ fm, whereas the chiral symmetry is likely recovered for the clover at $a= 0.06$ fm, by virtue of the fact that its $a_0$ correlator, including the coupling to the $\eta \pi$ ghost state, seems
to coincide with that of the overlap fermion. These can be considered discretization error. However, it is not clear if
what we observe is a simple $\mathcal{O}(a^2)$ error. In view of the fact that the ratios are at $\sim 0.5$ for both 
the $a = 0.12$ fm and $a = 0.09$ fm cases and they jump to unity at $a = 0.06$ fm, the possibility that chiral symmetry restoration may set in abruptly near the latter lattice spacing should be considered.
\begin{figure} [htb]
      \includegraphics[width=0.9\linewidth]{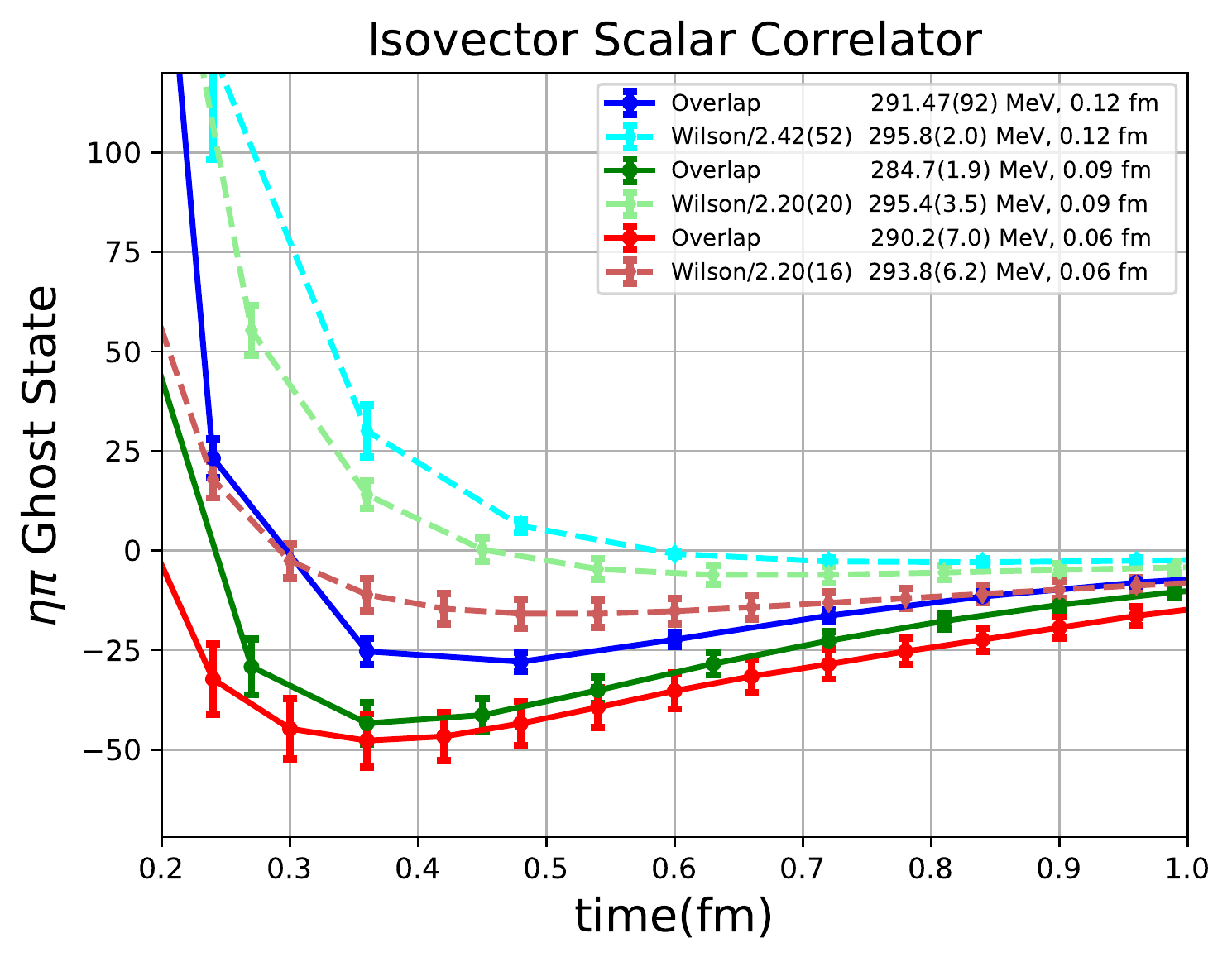}
    \caption{The isovector scalar meson ($a_0$) correlator with quenched
      configurations for Wilson and overlap fermions for a valence pion mass of
      $\sim 290$ MeV at three lattice spacings are plotted for comparison.}
  \label{fig:ghost}
\end{figure}

\begin{figure} [hb]
      \includegraphics[width=0.7\linewidth]{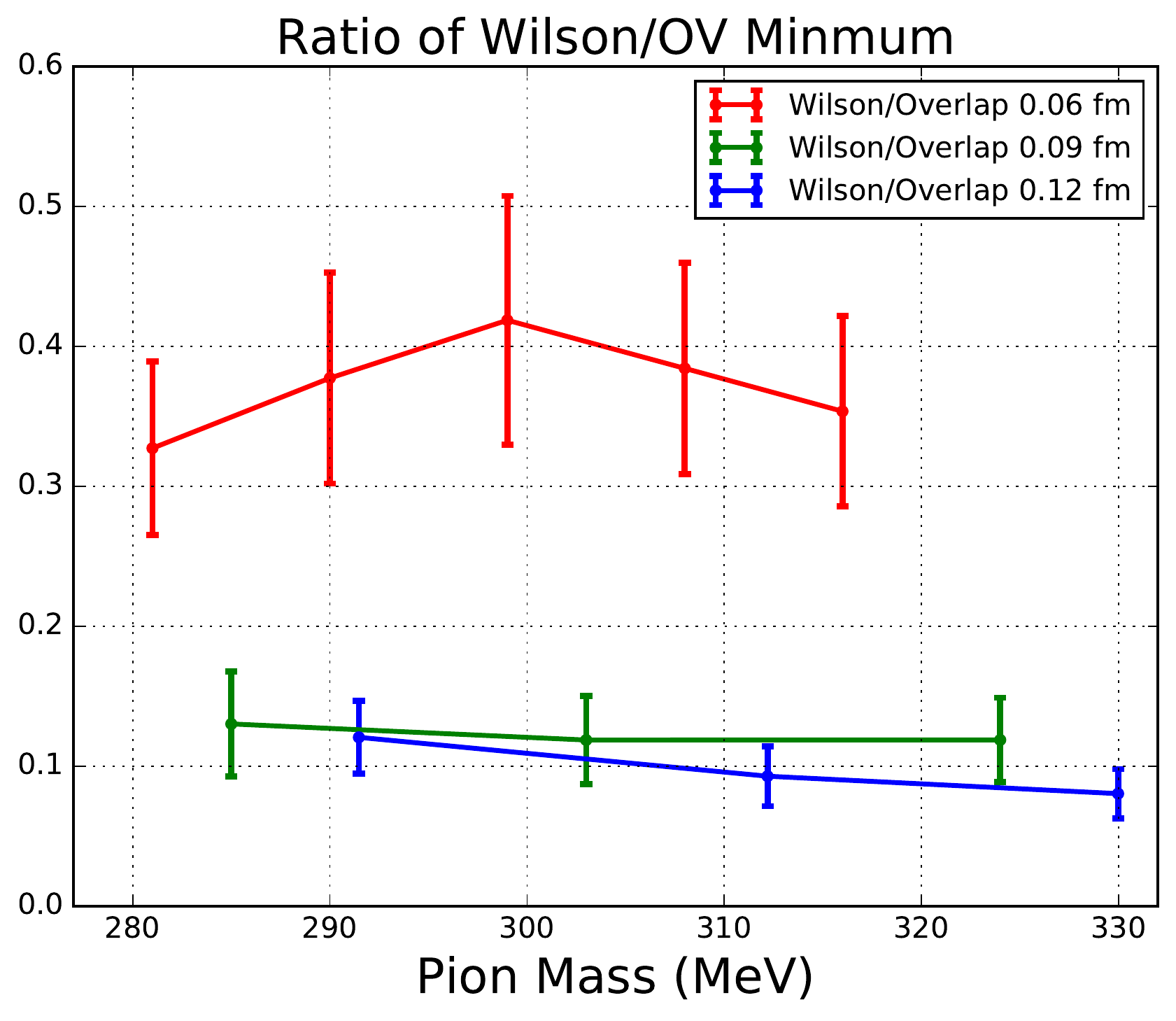}
    \caption{The ratios of the minima of the quenched $a_0$ correlator of the Wilson fermion to that of overlap fermion for a range of pion masses
  at  three lattice spacings. }
  \label{fig:min_compare}
\end{figure}

    We have also compared Wilson and overlap fermions to explore the discretization errors
related to chiral dynamics. Plotted in Fig.~\ref{fig:ghost} are the 
$a_0$ correlators from the Wilson and overlap fermions (without HYP smearing) on the same three quenched Wilson-gauge lattices with a pion mass around 290 MeV. As in the case of the above clover/overlap comparison, we
divide the Wilson $a_0$ correlator by a factor which is the ratio of the Wilson to overlap
correlator values at large time separation with the $a_0$ mass at $\sim 1.7$ GeV.
The factors are $2.43(52), 2.20(20), 2.20(16)$ for the lattices with spacings $0.12,
0.09$ and 0.06 fm, respectively. To put results from all three lattices in the same figure, we have multiplied the
$N_S^3$ factor from Eq.~(\ref{G-S}) to scale out the volume dependence  as far as the ghost state is concerned. 
We see from Fig.~\ref{fig:ghost} that the ghost states in the Wilson fermions are much shallower than
those the overlap. They are shallower than those of the clover fermion in Fig.~\ref{fig:ghost_clo/ov}. When the ratios
of the minima are plotted in Fig.~\ref{fig:min_compare} for the pion mass range of 280 to 330 MeV, we see that
the Wilson minima is only $\sim$ 10\% of that of the overlap for the $a = 0.12$ and 0.09 fm cases. At $a = 0.06$ frm,
the ratio is still only $\sim$ 40\%. This clearly demonstrates that Wilson fermion at these
lattice spacings, due to a lack of chiral symmetry, does not produce the full
$Z$-graphs in the hairpin diagram, as does the overlap fermion. Its coupling to the $\eta \pi$ ghost is still weaker 
than that of the clover. This is to be expected since Wilson has $\mathcal{O}(a)$
error, whereas clover has $\mathcal{O}(a^2)$ error and is chirally improved from the Wilson fermion.

How is the above observation of the ghost state in the $a_0$ channel related to
the fact that the Roper state is $\sim 300$ MeV lower than that of the clover
fermion that we studied in Sec.~III? As we mentioned in the Introduction: the
finding in the extensive dynamical coupled-channel (DCC)
study~\cite{JuliaDiaz:2007kz,Kamano:2010ud,Suzuki:2009nj}, the need of higher
Fock space in the electroexcitation experiment~\cite{Aznauryan:2011qj}, and the
pattern of parity reversal of the excited $N, \Delta$ in contrast to those of
$\Lambda$, all lead to the suggestion that there is a large higher Fock space 
component in the Roper state. It is shown explicitly in a chiral constituent quark model that the
flavor-spin interaction, due to the pseudoscalar meson exchange between the
quarks, is responsible for the parity pattern in $N, \Delta$, and $\Lambda$ and
the lowering of Roper from the $SU(6)$ quark model with the color-spin
interaction~\cite{Glozman:1997ag}.

\begin{figure}[ht]
  \centering
  \includegraphics[width=1.0\linewidth]{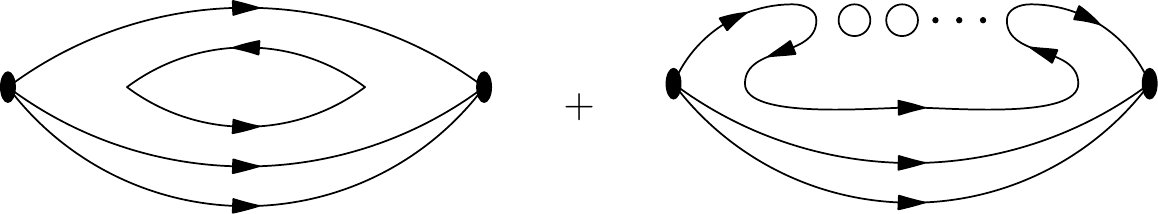}
  \caption{Quark loop contribution to the $\pi N$ and $\eta N$ intermediate states from a three-quark nucleon
  interpolation field.}
  \label{fig:eta-N}
\end{figure}

For the baryons, there are quark loop diagrams leading to $\pi N$ and $\eta N$
states which are depicted in Fig.~\ref{fig:eta-N}. They are analogous to the
$\eta\pi$ state in the $a_0$ correlator in Fig.~\ref{fig:eta-pi}.

In order to study the impact of the higher Fock-space components on the Roper,
lattice calculations with clover fermions at $m_{\pi} = 156$ MeV are carried
out with the inclusion of the $N\pi$ and $N\sigma$ type 5-quark
(i.e., $qqqq\bar{q}$) interpolation fields in addition to the 3-quark
interpolation field. Even though the expected $N\pi\pi$ and $N\pi$ scattering
states are observed, no additional state is found below 1.65 GeV
~\cite{Lang:2016hnn,Padmanath:2017oya}. It is concluded that the Roper does not
come down from the value calculated with the 3-quark interpolation field alone
in this $\pi N$ scattering calculation. A similar conclusion is reached with
mixed 3-quark and 5-quark interpolation fields for the clover fermion at
$m_{\pi} = 411$ MeV~\cite{Kiratidis:2016hda}.
\begin{figure}[ht]
  \centering
  \includegraphics[width=0.47\linewidth]{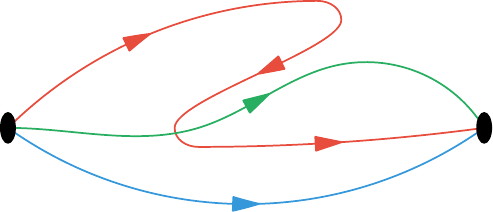}
  \hspace{0.2cm}
   \includegraphics[width=0.47\linewidth]{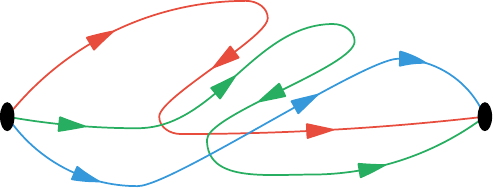}
  \caption{$Z$-graphs which lead to $\pi N$ and $\pi\pi N$ intermediate states
    from a three-quark nucleon interpolation field.}
  \label{fig:Zgraph-N}
\end{figure} 

To understand the results of these $\pi N$ scattering calculations, we point
out that there are other classes of diagrams that will lead to higher Fock
space components. These are depicted in Fig.~\ref{fig:Zgraph-N} as
$Z$-graphs. This can happen when there is more than one valence quark (or more
than one valence antiquark) in the interpolation field, such as for baryons. The 
back-bending of the valence quark in the Z-graph results in an antiquark 
propagating forward in time. It can combine with another forward propagating valence
quark to form a meson propagating between the two valence quarks as depicted in Fig.~\ref{fig:Zgraph-N}. 
It is worthwhile to note that this mechanism does not exist for the non-singlet mesons with $\bar{q}\Gamma q$-type
interpolation fields~\cite{Isgur:1999ic}. This is because the forward propagating 
antiquark from the Z-graph due to the valence quark and the other forward propagating
antiquark from the interpolation field do not form a meson. To do so, it would require one quark
and one antiquark. Thus additional $Z$-graphs in the
baryon can be cast effectively as the origin of meson exchanges between the
valence quarks in the chiral quark model~\cite{Liu:1999kq}. It is the
flavor-spin interaction from the Goldstone boson exchange that produces the
parity-reversal pattern in $N$ and $\Delta$ and lowers the Roper state to its
experimental value in the chiral constituent quark
model~\cite{Glozman:1997ag}. Including a 5-quark interpolation field will
entail coupling to higher Fock-space components, as illustrated in
Fig.~\ref{fig:eta-N}, but not through the $Z$-graphs in
Fig.~\ref{fig:Zgraph-N}, if the fermion action does not have a tendency to
generate $Z$-graphs dynamically at the given lattice spacing. The lattice calculations in this study, the
$\pi N$ scattering
calculations~\cite{Lang:2016hnn,Padmanath:2017oya,Kiratidis:2016hda}, the Roper
calculation in DCC~\cite{JuliaDiaz:2007kz,Kamano:2010ud,Suzuki:2009nj}, and the
chiral quark model~\cite{Glozman:1997ag}, all together hint strongly that it is the
$Z$-graphs in Fig.~\ref{fig:Zgraph-N}, innate to chiral dynamics, that are
responsible for lowering the Roper state and resulting in the parity reversal
in $N$ and $\Delta$ excited states. Therefore, only a lattice action with a
good chiral symmetry on the lattice will be able to capture the correct
dynamics of the Roper resonance and aid in the correct extraction of its energy
level. Using the overlap fermion action, which has exact chiral symmetry at 
finite lattice spacing, we are able to extract the energy of the Roper
resonance on fairly coarse lattices. On the other hand, at $a = 0.06$ fm or smaller lattice
spacing, the clover fermion may become chiral, in which case the  calculated Roper state should come
down to its experimental value.

\section{V. Summary and Conclusion}

We have calculated the Roper state on the lattice from the valence overlap
fermions with both the Sequential Empirical Bayes (SEB) method and the
variational method with large smeared sources and found consistent results. The
calculations were carried out on the $2+1$-flavor domain-wall fermion gauge
configurations on a $24^3 \times 64$ lattice with the lattice spacing $a =
0.114$ fm and light sea quark mass which corresponds to a pion mass of 330
MeV. The chirally extrapolated Roper mass is consistent with the experimental
result of N(1440).

The Roper masses for pion masses in the range 260--570 MeV are consistently
lower than those from the calculation with Wilson-type fermions by $\sim 300 -
600$ MeV. To further check the fitting algorithms, we applied both SEB and
variational approaches to the $2+1$-flavor clover fermion configurations on an
anisotropic $24^3 \times 128$ lattice with comparable spatial lattice spacing
and pion mass, i.e., with $a_s= 0.123$ fm, $\xi=3.5$, and $m_{\pi} = 390$
MeV. Again, we find that the SEB and variational calculations agree when a
large smearing-sized interpolation field is included in the variation
calculation.  The Roper mass on this clover lattice is $\approx 280$ MeV higher
than the value found on the previous lattice with overlap fermions at the same
pion mass.  It is clear that the difference is not due to the fitting
algorithm. 

To understand the definite difference between the two fermion
actions, we examine the coupling of single-hadron interpolating fields to two
hadron states. This explores the idea that the three-quark core couples to the
higher Fock space $N \pi, N \eta$ and $N \pi\pi$ states which brings down the
Roper mass from the quark model's uncoupled-three-quark-state value.  To this
end, we studied the would-be-$\eta\pi$ ghost state with the isovector scalar
$\overline{\psi}\psi$ interpolation field in the quenched approximation. In
this channel, it is well known that the would-be-$\eta\pi$ ghost state
dominates at pion masses lower than $\approx 600$ MeV for the overlap
fermion. This indicates that the one-hadron $\overline{\psi}\psi$ interpolation
field will couple strongly to the physical $\eta\pi$ and $\eta^\prime\pi$
two-hadron states in the dynamical fermion setting. 

We compare the $a_0$ correlators between the clover fermion and overlap fermion, both with HYP smearing. 
It is found that while the $\eta\pi$ ghost state shows up prominently with the overlap fermion on 
all three lattices at lattice spacings $a = 0.12, 0.09$ and 0.06 fm, the coupling to the ghost state
are not as strong in the clover cases for $a = 0.12, 0.09$ fm; the minima of their $a_0$ correlator which 
characterize the strength of the coupling to the $\eta\pi$ in the hairpin diagram are
only half as strong as those of their corresponding overlap case. On the other hand, the $a_0$ correlators
of the clover and overlap coincide within errors at $a = 0.06$ fm. We take this as an indication that
chiral symmetry is restored for the clover fermion at this lattice spacing. It is not clear, at this stage,
if this is due to the simple $\mathcal{O}(a^2)$ discretation error, or perhaps the chiral symmetry sets in abruptly near 
$a = 0.06$ fm. In any case, one would expect that the Roper will come down to the experimental value with the clover fermion at $a \le 0.06$ fm. We also compared Wilson and overlap fermions. The minima of the $a_0$ Wilson correlators
are much shallower than those of the overlap for the above three lattices, reflecting the large $\mathcal{O}(a)$ error
of the Wilson fermion. That the Wilson fermion is worse than the clover as far as the hairpin $Z$-graph is
concerned is consistent with the fact that clover is chirally improved from the Wilson fermion.

We surmise that the different lattice results, described in this work, for the
Roper state from the Wilson-type fermion versus overlap fermion and their associated
chiral behaviors for the quenched ghost would-be-$\eta\pi$ state supports a
resolution of the ``Roper puzzle'' due to $Z$-graph type chiral dynamics. This
entails coupling to higher Fock-space components (e.g.\ $N\pi$, $N\pi\pi$
states) to induce the effective flavor-spin interaction between quarks
prescribed in the chiral quark model, resulting in the parity-reversal pattern
as observed in the excited states of $N, \Delta$ and $\Lambda$.  This work is
also consistent with the conclusion about the existence of higher Fock-space
components in the Roper when experimental information on the nucleon to Roper
transition form factors is examined~\cite{Burkert:2017djo}.

{In this study, we concentrate on comparing the Roper state calculated from the 3-quark interpolation field and try to reconcile the difference between the results from the clover fermion and the overlap fermion. Since the width of the Roper is small compared to its mass, we believe that the mass calculated from the 3-quark interpolation field should be close to the resonance mass from the $\pi N$ scattering, much like the case of the $\rho$ resonance from the $\pi\pi$ scattering. In the future, when the pion mass is close to the physical one, it would be important to carry out a lattice calculation which includes the $N\pi$ and $N\pi\pi$ channels to locate the pole position of the Roper and its width from the phase shift.}

\section{Acknowledgments}
We thank the RBC and UKQCD Collaborations for providing their DWF
gauge configurations. This work is supported in part by the U.S. DOE
Grant No. DE-SC0013065  and DOE Grant No. DE-AC05-06OR23177 which is within 
the framework of the TMD Topical Collaboration. A. A. is supported by
National Science Foundation CAREER Grant No. PHY-1151648. A. A. and F. L. are supported 
by U.S. DOE Grant No. DE-FG02-95ER40907. 
A.L. thanks the Institute for Nuclear Theory at the University of Washington for its kind hospitality 
and stimulating research environment. His research was supported in part by the INT's 
U.S. DOE grant No. DE-FG02-00ER41132.   
This research used resources of the Oak Ridge
Leadership Computing Facility at the Oak Ridge National Laboratory,
which is supported by the Office of Science of the U.S. Department
of Energy under Contract No. DE-AC05-00OR22725. This work used Stampede
time under the Extreme Science and Engineering Discovery Environment
(XSEDE), which is supported by National Science Foundation Grant No.
ACI-1053575. We also thank the National Energy Research Scientific
Computing Center (NERSC) for providing HPC resources that have contributed
to the research results reported within this paper. We acknowledge
the facilities of the USQCD Collaboration used for this research in
part, which are funded by the Office of Science of the U.S. Department
of Energy.

\bibliography{roper}
\end{document}